\documentclass[a4paper]{article}

\usepackage{a4wide}
\usepackage{authblk}
\usepackage{latexsym}
\usepackage{bm}
\usepackage[english]{babel}
\usepackage{graphicx,subfigure,colortbl}
\usepackage{algorithm}
\usepackage{algpseudocode}
\usepackage{epstopdf}
	\graphicspath{{./}{figures/}}
\usepackage[colorlinks=true,linkcolor=blue,citecolor=blue]{hyperref}
\usepackage{xcolor}
\usepackage{amsfonts,amsmath,amsthm,mathtools,bbm,cool}
\usepackage{tikz}
\usetikzlibrary{arrows.meta, positioning, calc}

\tikzset{
    pics/agent/.style={
        code={
            \begin{scope}[shift={(0,-0.85)}]

            \shade[
                ball color=blue!20,
                draw=blue!65!black,
                line width=0.4pt
            ]
                (0,2.00) circle (0.28);

            \path[
                fill=blue!20,
                draw=cyan!65!black,
                line width=0.4pt
            ]
                (-0.11,1.76)
                -- (-0.10,1.59)
                -- (0.10,1.59)
                -- (0.11,1.76)
                -- cycle;

            \path[
                fill=blue!20,
                draw=blue!65!black,
                line width=0.4pt
            ]
                (-0.10,1.62)
                .. controls (-0.35,1.60) and (-0.47,1.45) ..
                (-0.44,1.18)
                -- (-0.34,0.55)
                -- (0.34,0.55)
                -- (0.44,1.18)
                .. controls (0.47,1.45) and (0.35,1.60) ..
                (0.10,1.62)
                -- cycle;

            \path[
                fill=blue!20,
                draw=cyan!65!black,
                line width=0.4pt
            ]
                (-0.34,1.48)
                .. controls (-0.53,1.30) and (-0.58,1.00) ..
                (-0.61,0.61)
                .. controls (-0.62,0.53) and (-0.56,0.48) ..
                (-0.50,0.51)
                -- (-0.38,1.20)
                -- cycle;

            \path[
                fill=blue!20,
                draw=cyan!65!black,
                line width=0.4pt
            ]
                (0.34,1.48)
                .. controls (0.53,1.30) and (0.58,1.00) ..
                (0.61,0.61)
                .. controls (0.62,0.53) and (0.56,0.48) ..
                (0.50,0.51)
                -- (0.38,1.20)
                -- cycle;

            \path[
                fill=blue!20,
                draw=blue!65!black,
                line width=0.4pt
            ]
                (-0.31,0.57)
                -- (-0.04,0.57)
                -- (-0.10,-0.46)
                .. controls (-0.11,-0.54) and (-0.18,-0.57) ..
                (-0.29,-0.56)
                -- (-0.37,-0.54)
                -- cycle;

            \path[
                fill=blue!20,
                draw=blue!65!black,
                line width=0.4pt
            ]
                (0.04,0.57)
                -- (0.31,0.57)
                -- (0.37,-0.54)
                -- (0.29,-0.56)
                .. controls (0.18,-0.57) and (0.11,-0.54) ..
                (0.10,-0.46)
                -- cycle;

            \end{scope}
        }
    }
}

\tikzset{
    pics/agent2/.style={
        code={
            \begin{scope}[shift={(0,-0.85)}]

            \shade[
                ball color=red!30,
                draw=red!65!black,
                line width=0.4pt
            ]
                (0,2.00) circle (0.28);

            \path[
                fill=red!40,
                draw=red!65!black,
                line width=0.4pt
            ]
                (-0.11,1.76)
                -- (-0.10,1.59)
                -- (0.10,1.59)
                -- (0.11,1.76)
                -- cycle;

            \path[
                fill=red!35,
                draw=red!65!black,
                line width=0.4pt
            ]
                (-0.10,1.62)
                .. controls (-0.35,1.60) and (-0.47,1.45) ..
                (-0.44,1.18)
                -- (-0.34,0.55)
                -- (0.34,0.55)
                -- (0.44,1.18)
                .. controls (0.47,1.45) and (0.35,1.60) ..
                (0.10,1.62)
                -- cycle;

            \path[
                fill=red!30,
                draw=red!65!black,
                line width=0.4pt
            ]
                (-0.34,1.48)
                .. controls (-0.53,1.30) and (-0.58,1.00) ..
                (-0.61,0.61)
                .. controls (-0.62,0.53) and (-0.56,0.48) ..
                (-0.50,0.51)
                -- (-0.38,1.20)
                -- cycle;

            \path[
                fill=red!30,
                draw=red!65!black,
                line width=0.4pt
            ]
                (0.34,1.48)
                .. controls (0.53,1.30) and (0.58,1.00) ..
                (0.61,0.61)
                .. controls (0.62,0.53) and (0.56,0.48) ..
                (0.50,0.51)
                -- (0.38,1.20)
                -- cycle;

            \path[
                fill=red!40,
                draw=red!65!black,
                line width=0.4pt
            ]
                (-0.31,0.57)
                -- (-0.04,0.57)
                -- (-0.10,-0.46)
                .. controls (-0.11,-0.54) and (-0.18,-0.57) ..
                (-0.29,-0.56)
                -- (-0.37,-0.54)
                -- cycle;

            \path[
                fill=red!40,
                draw=red!65!black,
                line width=0.4pt
            ]
                (0.04,0.57)
                -- (0.31,0.57)
                -- (0.37,-0.54)
                -- (0.29,-0.56)
                .. controls (0.18,-0.57) and (0.11,-0.54) ..
                (0.10,-0.46)
                -- cycle;

            \end{scope}
        }
    }
}

\tikzset{
    pics/agent3/.style={
        code={
            \begin{scope}[shift={(0,-0.85)}]

            \shade[
                ball color=green!30,
                draw=green!65!black,
                line width=0.4pt
            ]
                (0,2.00) circle (0.28);

            \path[
                fill=green!40,
                draw=green!65!black,
                line width=0.4pt
            ]
                (-0.11,1.76)
                -- (-0.10,1.59)
                -- (0.10,1.59)
                -- (0.11,1.76)
                -- cycle;

            \path[
                fill=green!35,
                draw=green!65!black,
                line width=0.4pt
            ]
                (-0.10,1.62)
                .. controls (-0.35,1.60) and (-0.47,1.45) ..
                (-0.44,1.18)
                -- (-0.34,0.55)
                -- (0.34,0.55)
                -- (0.44,1.18)
                .. controls (0.47,1.45) and (0.35,1.60) ..
                (0.10,1.62)
                -- cycle;

            \path[
                fill=green!30,
                draw=green!65!black,
                line width=0.4pt
            ]
                (-0.34,1.48)
                .. controls (-0.53,1.30) and (-0.58,1.00) ..
                (-0.61,0.61)
                .. controls (-0.62,0.53) and (-0.56,0.48) ..
                (-0.50,0.51)
                -- (-0.38,1.20)
                -- cycle;

            \path[
                fill=green!30,
                draw=cyan!65!black,
                line width=0.4pt
            ]
                (0.34,1.48)
                .. controls (0.53,1.30) and (0.58,1.00) ..
                (0.61,0.61)
                .. controls (0.62,0.53) and (0.56,0.48) ..
                (0.50,0.51)
                -- (0.38,1.20)
                -- cycle;

            \path[
                fill=green!40,
                draw=green!65!black,
                line width=0.4pt
            ]
                (-0.31,0.57)
                -- (-0.04,0.57)
                -- (-0.10,-0.46)
                .. controls (-0.11,-0.54) and (-0.18,-0.57) ..
                (-0.29,-0.56)
                -- (-0.37,-0.54)
                -- cycle;

            \path[
                fill=green!40,
                draw=green!65!black,
                line width=0.4pt
            ]
                (0.04,0.57)
                -- (0.31,0.57)
                -- (0.37,-0.54)
                -- (0.29,-0.56)
                .. controls (0.18,-0.57) and (0.11,-0.54) ..
                (0.10,-0.46)
                -- cycle;

            \end{scope}
        }
    }
}

\newtheorem{theorem}{Theorem}[section]

\theoremstyle{remark}\newtheorem{remark}[theorem]{Remark}

\newcommand{\be}{\begin{equation}}
\newcommand{\ee}{\end{equation}}
\newcommand{\eps}{\varepsilon}
\newcommand{\eq}{\mathrm{eq}}
\newcommand{\init}{\mathrm{in}}

\newcommand{\R}{\mathbb{R}}

\newcommand{\dd}{\mathrm{d}}

\newcommand{\norm}[1]{\left\|#1\right\|}

\allowdisplaybreaks

\begin{document}
\title{Collective impact of opinion formation\\on agent-based epidemic dynamics}

\author[1]{Andrea Bondesan\thanks{\texttt{andrea.bondesan@unipr.it}}} 

\author[2]{Riccardo Vecchio\thanks{\texttt{riccardo.vecchio@unipv.it}}}

\author[2]{Chiara Barbati\thanks{\texttt{chiara.barbati@universitadipavia.it}}}

\author[3]{Jonathan Franceschi\thanks{\texttt{jonathan.franceschi@unife.it}}}

\author[2]{Anna Odone\thanks{\texttt{anna.odone@unipv.it}}}

\author[4]{Mattia Zanella \thanks{\texttt{mattia.zanella@unipv.it}}}

\affil[1]{Department of Mathematical, Physical and Computer Sciences, University of Parma, Italy \vspace{1mm}}

\affil[2]{Department of Public Health, Experimental and Forensic Medicine, University of Pavia, Italy \vspace{1mm}}

\affil[3]{Department of Mathematics and Computer Sciences, University of Ferrara, Italy \vspace{2mm}}

\affil[4]{Department of Mathematics ``F. Casorati'', University of Pavia, Italy}

\maketitle

\begin{abstract}
Epidemic spread and collective opinion formation are tightly coupled processes: individual beliefs about the severity of an outbreak and the value of protective measures shape behavior, while the evolving epidemic itself reshapes public sentiment through risk perception. We propose an agent-based framework that couples opinion dynamics with a compartmental epidemic model, in which each agent carries both a health status and a continuous opinion variable describing adherence to protective measures. Opinion evolution is driven by binary interactions among agents and by an exogenous background field that reflects the current state of the epidemic, while the disease transmission rate depends explicitly on the opinions of interacting agents. Using methods of statistical mechanics, we characterize the resulting macroscopic system for the compartmental mass fractions and mean opinions in both far-from-equilibrium and close-to-equilibrium regimes. The calibration shows that incorporating behavioral opinion dynamics substantially improves the interpretation of epidemic evolution. In particular, the inferred risk perception function evolves consistently with the observed epidemic burden, indicating that behavioral adaptation can explain part of the temporal variation that would otherwise be absorbed into time-dependent transmission parameters. Our results illustrate how integrating opinion heterogeneity in kinetic epidemic models offers quantitative insights into the feedback between behavior and disease spread.
\medskip


\end{abstract}

\tableofcontents

\section{Introduction}
\label{sect:intro}

Epidemic dynamics and opinion formation appear to be strictly coupled dynamics. Indeed, individuals' opinions regarding the severity of an infectious disease outbreak and the effectiveness of control measures influence their behavioral choices, which in turn affect transmission patterns. At the same time, the evolution of the epidemic acts as a driver of opinion changes through the perception of risk and social interactions \cite{Funk_PNAS,PhysRevE.90.012808,Liu_IEEE}. This bidirectional feedback between agent-based opinion dynamics and infection spread motivates the development of new models and methods to understand and control the impact of epidemic processes. In more detail, during the early stages of the COVID-19 pandemic, in the absence of effective treatments, recent experimental evidence showed that the spread of an infectious disease is strongly shaped by how widely populations adhere to non-pharmaceutical interventions (NPIs), see, e.g., \cite{Tu,Zhang_etal}, making epidemic dynamics linked to opinion formation phenomena \cite{DC}. In recent years, among the many aspects of opinion dynamics, vaccine hesitancy has received particular attention, owing to its far-reaching consequences for both individual and public health \cite{Dub,LoM,ODONE_SIGNORELLI,VBA}. In this direction, a variety of mathematical frameworks have been proposed to capture this phenomenon, see e.g. \cite{Ber_etal,BTZ,BDM} and the references therein. Notably, vaccine hesitancy is tightly connected to the spread of misinformation \cite{FPBGB,MR4493790,garett,odone20}.

Classical compartmental models, such as the susceptible–infectious–removed (SIR) and susceptible–exposed–infectious–removed (SEIR) frameworks, provide a well-established representation of infectious disease transmission by describing population flows across epidemiological states \cite{bjon20}. However, transmission is also influenced by dynamic and context-dependent factors, including individual behavior, risk perception, public-health communication and adherence to protective measures \cite{mossong}. As behavioral adaptation is not usually represented explicitly, the interactions and temporal evolution of these factors may not be fully captured by standard compartmental models \cite{mossong}. 

Several research efforts concentrated in trying to understand how the interplay between opinion dynamics and epidemic spreading can be captured within a unified mathematical framework. Several approaches have coupled compartmental epidemic models with imitation dynamics mechanisms, allowing infection and behavior dynamics to evolve jointly \cite{DMLM,PCAPM,ZM}. Other contributions have adopted a multi-agent perspective or kinetic perspective, describing the joint evolution of health status and opinion as a microscopic process, see the seminal work \cite{Tos}, from which macroscopic epidemic-opinion dynamics can be derived \cite{GFPS,PSHZ,Zan}. Network-based models have also been employed to account for the role of social structure in shaping both the diffusion of information and the transmission of infectious diseases \cite{A_etal,ACDPR,BonBorFon,daSilva19}. Despite these efforts, a comprehensive understanding of how opinion heterogeneity and its evolution affect the qualitative and quantitative features of an epidemic outbreak, such as peak prevalence or incidence, epidemic patterns, remains an open and active area of research, see e.g. \cite{Block:2020aa,daSilva19,DG,SPG}.

A natural mathematical setting for going beyond purely individual-based descriptions of such many-agent systems is offered by kinetic equations, in which the collective behavior of large interacting populations is studied through the tools of statistical mechanics \cite{BerLorSenTos,BerParTos,MR4770477,CPS,DMLT,FHT,MT,PT13}. Within this setting, the study of consensus formation and opinion dynamics has received substantial attention in recent years \cite{BonBor,BonBorFon,CT,DurFraWolZan,DMPW,FraMedZan}, and kinetic descriptions have proven especially well suited to coupling microscopic opinion exchange with epidemiological compartments, since they naturally accommodate heterogeneous, opinion-dependent transmission mechanisms without requiring an a priori closure of the resulting macroscopic system.

In this work, we build on this kinetic perspective to propose an agent-based model in which each individual is described jointly by an epidemiological compartment and a continuous opinion variable representing adherence to protective behavior. Opinion evolves through binary interactions between agents possibly belonging to different compartments and through interaction with a background information field whose target opinion is driven by the current state of the epidemic, reflecting the influence of media reporting and public health communication \cite{Tu}. The transmission rate between susceptible and infectious agents is in turn modulated by their respective opinions, closing the feedback loop between beliefs and disease spread. Resorting to a quasi-invariant interaction limit, we derive a reduced-complexity Fokker--Planck description of the opinion dynamics, whose equilibrium states can be computed explicitly as beta distributions, and we use this reduction to obtain closed macroscopic systems for the compartmental mass fractions and mean opinions, both far from and close to equilibrium. We then validate the resulting framework against real epidemiological and social media data from the first three months of the COVID-19 pandemic in Italy, using the model to both calibrate the transmission and risk perception parameters and to quantify, in a data-driven way, how representative a biased social media sample is of the sentiment of the wider population.

The rest of the paper is organized as follows. In Section~\ref{sec:model} we introduce the agent-based kinetic model coupling opinion dynamics with epidemic spread, and derive its quasi-invariant Fokker--Planck reduction together with the explicit equilibrium distributions. In Section~\ref{subsect:MF} we derive a reduced complexity model. Thanks to these results, in Section \ref{sec:macro} we discuss the derivation of the associated macroscopic systems for mass fractions and mean opinions, both in a far-from-equilibrium regime and, via a moment-closure argument, in a close-to-equilibrium regime. In Section~\ref{sec:numerics} we present the several illustrative numerical tests of the proposed model together with a calibration of the model against COVID-19 data from Italy, including the estimation of the risk perception function from a sample of social media interactions.

\section{Agent-based opinion dynamics and epidemic spread} \label{sec:model}

Epidemic dynamics and opinion formation are intrinsically coupled processes: individuals' opinions regarding the severity of an outbreak  influence their behavioral choices, which in turn affect transmission patterns. At the same time, the evolution of the epidemic acts as a driver of opinion changes through the perception of personal threat. This bidirectional feedback between agents' opinions and observable evolution of transmission motivates the development of new models capable of linking the emergence of collective opinion patterns with their impact on epidemic outcomes.

To this end, we consider a large system of agents in which each individual updates its opinion and may belong to different compartments. Seminal approaches have been proposed in \cite{BZ, BTZ, Zan}.

Resorting to classical epidemiological models based on a compartmentalization of large agent systems, we subdivide the population into the three epidemiologically relevant states as follows:  susceptible (S) agents -- that can contract the disease -- infectious (I) agents -- responsible for the spread of the disease -- and removed (R) agents -- that cannot spread the disease. Furthermore, each agent in the various compartments is characterized by an opinion variable $w \in \mathcal{I}$, where $\mathcal{I} = [-1,1]$, where $-1$ and $+1$ denote the two opposite beliefs regarding the protective behavior. In particular, the value $w = -1$ is associated with agents that do not believe in the necessity of protections (like wearing masks or reducing daily contacts), whereas $w = 1$ is associated with agents that are in complete agreement with protective behaviors and that are willing to adhere to  the protective measures. The continuous variable $w \in [-1,1]$ can be therefore intended as the propensity to adopt protective behaviors of a single agent.

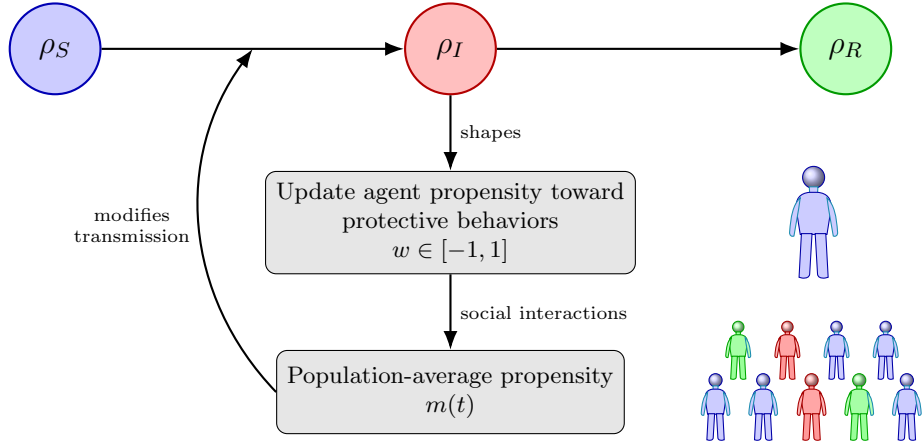
\begin{figure}[t]
\centering
\begin{tikzpicture}[
    compartment/.style={
        circle,
        draw,
        thick,
        minimum size=12mm,
        font=\large
    },
    opinion/.style={
        draw,
        rounded corners,
        align=center,
        minimum width=3.7cm,
        minimum height=1.1cm,
        inner sep=4pt,
        font=\small
    },
    transition/.style={
        -{Latex[length=2.5mm]},
        thick
    },
    feedback/.style={
        -{Latex[length=2.5mm]},
        thick,
        dashed
    },
    every edge quotes/.style={
        font=\small,
        align=center
    }
]

\node[compartment,fill=blue!20,
        draw=blue!70!black] (S) {$\rho_S$};
\node[compartment, right=4cm of S, fill=red!25,
        draw=red!70!black] (I) {$\rho_I$};
\node[compartment, right=4cm of I, fill=green!25,
        draw=green!60!black] (R) {$\rho_R$};

\draw[transition]
    (S) -- node[above, align=center]
    {}
    (I);

\draw[transition]
    (I) -- node[above]
    {}
    (R);

\node[opinion, below=1cm of I,fill=black!10] (individual)
    {Update agent propensity toward\\
     protective behaviors\\
     $w\in[-1,1]$};
     
\coordinate[right=2.35cm of individual] (agent);

\pic[scale=0.52]
    at (agent) {agent};

\node[opinion, below=1cm of individual,fill=black!10] (mean)
    {Population-average propensity\\
     $m(t)$};
     
\coordinate[right=2.15cm of mean]
    (populationCentre);

\begin{scope}[shift={(populationCentre)}]

    \pic[scale=0.27] at (-0.65,0.55) {agent3};
    \pic[scale=0.27] at ( 0.00,0.55) {agent2};
    \pic[scale=0.27] at ( 0.65,0.55) {agent};
    \pic[scale=0.27] at ( 1.3,0.55) {agent};

    \pic[scale=0.31] at (-0.95,-0.20) {agent};
    \pic[scale=0.31] at (-0.32,-0.20) {agent};
    \pic[scale=0.31] at ( 0.32,-0.20) {agent2};
    \pic[scale=0.31] at ( 0.95,-0.20) {agent3};
    \pic[scale=0.31] at ( 1.58,-0.20) {agent};

\end{scope}

\draw[transition]
    (I) -- node[right, font=\scriptsize]
    {shapes}
    (individual);

\draw[transition]
    (individual) to node[right, font=\scriptsize]
    {social interactions}
    (mean);

\draw[transition]
    (mean.west) to[bend left=40]
    node[left, font=\scriptsize, align=center]
    {modifies\\transmission}
    ($(S)!0.5!(I)$);

\end{tikzpicture}

\caption{\textbf{Opinion-dependent dynamics.} The number of active cases shapes
risk perception, which influences individual propensities to adhere to
protective behaviors. Their population average \(m(t)\) modifies the
transmission rate from the susceptible to the infectious compartment.}
\label{fig:opinion-dependent-SIR}
\end{figure}

Based on this setting, we model the evolution of the system of compartments $\mathcal{C} = \{S,I,R\}$ through a vector distribution function $(f_J)_{J \in \mathcal{C}}$, where each $f_J = f_J(w,t)$ describes how agents with opinion $w \in \mathcal{I}$ are distributed over time $t \in \R_+$ in the corresponding compartment $J \in \mathcal{C}$. For each $f_J$ we introduce the macroscopic observables
\begin{equation} \label{eq:mass fractions}
    \rho_J(t) = \int_{\mathcal{I}} f_J(w,t) \dd w, \qquad \rho(t) = \sum_{J \in \mathcal{C}} \rho_J(t),
\end{equation}
defining respectively the number of agents in each compartment $J \in \mathcal{C}$, and the total number of the population. We also define the successive moments of order $\ell > 0$ of each distribution $f_J$ as
\begin{equation} \label{eq:successive moments}
    m_{J,\ell}(t) = \frac{1}{\rho_J(t)} \int_{\mathcal{I}} w^\ell f_J(w,t) \dd w.
\end{equation}
In particular, in the following we will use the shorthand notation $m_J(t) = m_{J,1}(t)$ to indicate the first order moments defining the compartmental mean opinions. These are given, together with the total mean opinion $m(t)$, by
\begin{equation} \label{eq:mean opinions}
    m_J(t) = \frac{1}{\rho_J(t)} \int_{\mathcal{I}} w f_J(w,t) \dd w, \qquad m(t) = \frac{1}{\rho(t)} \sum_{J \in \mathcal{C}} \rho_J(t) m_J(t).
\end{equation}
The quantity $m_J$ is therefore the average  propensity of the population of agents belonging to the compartment $J \in \mathcal C$. 
This formulation provides a unified kinetic framework in which the evolution of the epidemic and the collective opinion dynamics are treated as two interconnected processes, making it possible to investigate how changes in public perception influence epidemic outcomes and, conversely, how the epidemic itself reshapes collective beliefs.  We depict in Figure \ref{fig:opinion-dependent-SIR} the impact of opinion dynamics encapsulated in our kinetic model approach highlighting the epidemic evolution.  

In the next section, we couple an established model for the opinion evolution based on binary updates between agents. 

\subsection{Kinetic modeling of opinion interactions} \label{sect:2.1}

Each agent from a certain compartment $J \in \mathcal{C}$ modifies its opinion in two ways, by interacting with other agents (from the same $J = K$ or from different compartments $J \neq K$) and with a background environment whose opinion evolves depending on the state of the epidemic itself at each time $t \in \R_+$. 

Therefore, we assume that opinion formation is driven by binary interpersonal interactions, reflecting the idea that changes in individual beliefs arise primarily through pairwise exchanges between agents. From a public health perspective, these interactions represent everyday social contacts through which individuals influence each other's perception of the epidemic and their willingness to adopt protective behaviors. More precisely, an interaction between two individuals belonging to compartments $J,K \in \mathcal{C}$ and having respective opinions $(w,w_*)\in \mathcal{I}\times \mathcal{I}$ leads to the formation of a new opinion pair $(w',w_*')\in \mathcal{I}\times \mathcal{I}$ which is given by 
\begin{equation} \label{eq:bin_int1}
    \begin{split}
        w' &= w + \lambda P(w,w_*)(w_*-w) + D(w)\eta_J, \\[2mm]
        w_*' &= w_* + \lambda P(w_*,w)(w-w_*) + D(w_*) \eta_K,
    \end{split}
\end{equation}
where $P(w,w_*)\in [0,1]$ is a symmetric function $P(w,w_*) = P(w_*,w)$ modeling the compromise propensity, and $\lambda\in (0,1)$. In \eqref{eq:bin_int1} we then indicate with $D(w) \geq 0$ the local relevance of the self-thinking, mimicking possible deviations from the consensus dynamics, while $(\eta_J)_{J \in \mathcal{C}}$ denote i.i.d. random variables encoding the random self-thinking processes. In particular, throughout this work we will consider the function $D(w) = \sqrt{1-w^2}$ \cite{Tos}, which conveys the idea that an agent's opinion $w$ is more likely to change due to self-thinking when it is close to zero, while extreme opinions $w \sim \pm 1$ will unlikely change opinion due to self-thinking. Moreover, we assume that $\mathbb{E}[\eta_J] = 0$ and $\mathbb{E}[\eta_J^2] = \sigma_J^2 > 0$, $J \in \mathcal{C}$. We notice that the assumption of symmetry of the interaction function $P$ implies that the the post-interaction average opinion is conserved, i.e. $\mathbb{E}[w' + w'_*] = w + w_*$, and that the mean energy is dissipated when $\sigma_J^2 = 0$ for any $J \in \mathcal{C}$. We refer the interested reader to \cite{PTTZ} for a detailed discussion on this matter. We also point out that it is possible to bound the support of the random variables $(\eta_J)_{J \in \mathcal{C}}$ to guarantee that the post-interaction opinions defined in \eqref{eq:bin_int1} are such that $w',w'_* \in \mathcal{I}$, see \cite{Tos,Zan}. 

At the same time, each agent is exposed to a background information environment, whose message evolves according to the current state of the epidemic and models the effect of public health communication, media reporting, and the observed disease burden. Consequently, an agent belonging to compartment $J\in\mathcal C$ updates its opinion through the combined action of pairwise interactions with agents from any epidemiological compartment and the external information field. We characterize the influence of the current state of the epidemic on the opinion formation dynamics, measured in terms of the number of infected individuals $\rho_I(t) = \int_{\mathcal{I}} f_I(w,t) \dd w$, $t \in \R_+$, through interactions with an evolving background opinion distribution denoted by $g = g(z,t)$, with $z \in \mathcal{I}$, such that \cite{BTZ}
\begin{equation*}
    \int_{\mathcal{I}}g(z,t)\dd z = 1,\qquad \int_{\mathcal{I}}z g(z,t)\dd z = \mu(\rho_I(t)) \in \mathcal{I}, 
\end{equation*}
where $\mu = \mu(s)$ is assumed to be an increasing function $\mu:[0,1] \to \mathcal{I} \in C^1([0,1])$, satisfying the constraints $\mu(0) = -1$ and $\mu(1) = 1$. In particular, this function models the reaction of the population to the current perceived risk of the epidemic \cite{BTZ}, since the presence of many infectious agents $\rho_I \sim 1$ favors the assumption of protective behaviors $\mu \sim 1$ in an
effort to reduce the cases, while a reduction in the epidemic impact $\rho_I \sim 0$ pushes the population to take riskier decisions $\mu \sim -1$. The microscopic opinion updates are modeled as follows
\begin{equation} \label{eq:bin_int2}
\begin{split}
    w'' &= w + \lambda_B(z-w) + D(w)\eta_B,\\
    z^{\prime\prime} &= z,
    \end{split}
\end{equation}
where $\lambda_B \in (0,1)$ and $\eta_B$ is a centered random variable with finite variance $\mathbb{E}[\eta_B^2] = \sigma_B^2$. In particular, we assume that $\eta_B$ has the same distribution of the $(\eta_J)_{J \in \mathcal{C}}$ introduced in \eqref{eq:bin_int1}. 

The collective trends of the multi-agent system, where each agent updates its opinion following the binary scheme \eqref{eq:bin_int1} and the interaction with the current epidemiological state \eqref{eq:bin_int2}, are then determined by the Boltzmann-type model \cite{A_etal}
\begin{equation} \label{eq:kinetic}
\begin{split}
    \frac{\partial}{\partial t} f_J(w,t) = \sum_{K \in \mathcal{C}} Q_{JK}(f_J,f_K)(w,t) + Q^B(f_J,\rho_I), \qquad t \in \R_+,\; w \in \mathcal{I},\; J \in \mathcal C.
\end{split}
\end{equation}
In \eqref{eq:kinetic} the operators $Q_{JK}(f_J,f_K)$ accounts for all the binary exchanges \eqref{eq:bin_int1} of opinion between individuals and is defined as follows
\begin{equation} \label{eq:QC}
    Q_{JK}(f_J,f_K)(w,t) = \mathbb{E} \left[\int_{\mathcal{I}} \chi \left( \frac{1}{\mathcal{J}'} f_J(w',t) f_K(w'_*,t) - f_J(w,t) f_K(w_*,t) \right) \dd w_* \right], 
\end{equation}
where $(w',w'_*)$ is the pre-interaction opinion pair generating the post-interaction opinion pair $(w,w_*)$ according to the binary rules \eqref{eq:bin_int1}, while $\mathcal{J}'$ denotes the Jacobian of the transformation $(w',w'_*) \mapsto (w,w_*)$, and $\chi \geq 0$ characterizes the frequency of interactions \eqref{eq:bin_int1} between agents.

Similarly, the operator $Q_J^B(f_J,\rho_I)$, ${J \in \mathcal{C}}$, takes into account the opinion relaxation process expressed by \eqref{eq:bin_int2} and is defined in strong form as 
\begin{equation} \label{eq:QB}
    Q^B(f_J,\rho_I)(w,t) = \mathbb{E} \left[\int_{\mathcal{I}} \chi^B \left( \frac{1}{\mathcal{J}''}f_J(w'',t) g(z,t) - f_J(w,t) g(z,t) \right) \dd z \right], 
\end{equation}
where $w''$ is the pre-interaction opinion producing the post-interaction one $w$ according to the linear rules \eqref{eq:bin_int2}, $\mathcal{J}''$ is the Jacobian of the transformation $w'' \mapsto w$, and $\chi^B \geq 0$ encodes the frequency of interactions \eqref{eq:bin_int2} with the background opinion.  

\begin{remark}
On the one hand, the agent--agent interactions defined by \eqref{eq:bin_int1} describe the endogenous social dynamics of opinion formation: individuals modify their opinions through pairwise encounters with other individuals, possibly belonging to the same or to different epidemiological compartments. On the other hand, interaction with the epidemic's social perception defined by \eqref{eq:bin_int2} describe an exogenous or state-dependent mechanism through which an individual's opinion is modified by the current epidemiological situation. In accordance with \cite{PCAPM}, these two mechanisms may naturally operate on different time scales, characterized here by the interaction frequencies $\chi,\chi^B\geq0$. In particular, the regime
\begin{equation*}
    \frac{\chi^B}{\chi} \to 0, \qquad \chi,\chi^B \to +\infty,
\end{equation*}
corresponds to a time scale separation in which agent--agent interactions occur on a faster time scale than the response to the epidemic, while both mechanisms evolve on a faster time scale than the epidemiological dynamics. In this regime, opinions rapidly adapt through interpersonal interactions, whereas the influence of the evolving epidemic acts on a slower time scale. 

To comply with this scaling regime, throughout this work we assume that the interaction frequencies $\chi$ and $\chi^B$ are constant and given by
\begin{equation*}
    \chi=\frac{1}{\tau^2}, \qquad \chi^B=\frac{1}{\tau},
\end{equation*}
where the parameter $\tau > 0$ is measured in units of time. Hence, as $\tau \to 0$,
\begin{equation*}
    \chi \gg \chi^B \gg 1,
\end{equation*}
and the characteristic time scales associated with the agent--agent and agent--epidemic interactions are of order $\mathcal{O}(\tau^2)$ and $\mathcal{O}(\tau)$ respectively, while the epidemiological dynamics evolves on an order $\mathcal{O}(1)$ time scale. This scaling conveys the idea that interpersonal opinion exchanges occur much more frequently than interactions with the background sentiment induced by the epidemic. This is a reasonable modeling assumption, as the perception of the danger posed by an emerging disease may require a sufficiently long time to develop and induce behavioral awareness at the population level. Conversely, opinion exchanges between individuals are expected to occur more frequently, thereby motivating the faster time scale associated with the interactions \eqref{eq:bin_int1}.
\end{remark}

\subsubsection{Evolution of the observable quantities}

One can obtain the evolution of the main macroscopic observables (mass fractions, average compartmental opinions, etc.) by looking at the weak formulation of system \eqref{eq:kinetic}. Indeed, multiplying both sides of the equation by a suitable test function $\varphi = \varphi(w)$ and integrating over $w \in \mathcal{I}$, it is possible to rewrite \eqref{eq:kinetic} as
\begin{equation*}
    \begin{split}
        \frac{\dd}{\dd t} \int_{\mathcal{I}} \varphi(w) f_J(w,t)\dd w &= \frac{1}{\tau^2} \sum_{K \in \mathcal{C}} \mathbb{E}\left[\int_{\mathcal{I}\times \mathcal{I}} (\varphi(w') -\varphi(w)) f_J(w,t) f_K(w_*,t) \dd w\;\dd w_*\right] \\[4mm]
        & \qquad + \frac{1}{\tau} \mathbb{E} \left[\int_{\mathcal{I} \times \mathcal{I}} (\varphi(w'')-\varphi(w)) f_J(w,t) g(z,t) \dd w\; \dd z \right],
    \end{split}
\end{equation*}
for all $J\in \mathcal{C}$. First of all, choosing $\varphi(w) = 1$ we can infer that the mass fraction $\rho_J$ in each compartment $J \in \mathcal{C}$ is conserved over time, hence $\rho_J(t) = \rho_J(0)$ for any $t \in \R_+$ and also $\rho(t) = \rho(0) = 1$ without loss of generality. Furthermore, taking $\varphi(w) = w$ and summing over $J \in \mathcal{C}$ we can see that the operators \eqref{eq:QC} preserve total mean opinion $m(t)$, hence its evolution is solely dictated by the interactions with the background through operators \eqref{eq:QB}, leading to
\begin{equation} \label{eq:evolution of m}
    \frac{\dd}{\dd t} m(t) = -\frac{\lambda_B}{\tau} (m(t) - \mu(\rho_I)), \qquad t \in \R_+.
\end{equation}
In particular, when $\tau \to +\infty$ the alignment to the background opinion distribution is negligible and the total mean opinion of the population is conserved over time, i.e. $m(t) = m(0)$ for any $t \in \R_+$, whereas for finite $\tau > 0$ the mean opinion $m(t)$ converges asymptotically to the (constant, since $\rho_I$ is conserved) background mean opinion $\mu(\rho_I)$  
\begin{equation*}
    m(t) = m(0)e^{-\frac{\lambda_B}{\tau}t} + \mu(\rho_I)\left(1-e^{-\frac{\lambda_B}{\tau}t}\right).
\end{equation*}
More in general, it is a central topic in classical kinetic theory of gases to derive the explicit equations that drive the evolution of the relevant macroscopic quantities \cite{PT13}. Since determining these equations for the higher order moments $m_{J,\ell}$, $\ell \geq 2$, is a difficult task, the typical approach is to look at close-to-equilibrium dynamics of system \eqref{eq:kinetic}. One would like to transfer the same strategy to the study of multi-agent kinetic equations, but in this case it becomes much harder to identify their long-time equilibrium distributions, due to the complexity of generalized microscopic interactions like \eqref{eq:bin_int1}--\eqref{eq:bin_int2}. A possible strategy to overcome such problem consists in reducing the Boltzmann operators \eqref{eq:QC}--\eqref{eq:QB} to simplified drift--diffusion operators of Fokker--Planck type, for which the study of asymptotic properties becomes significantly easier.


\subsubsection{Reduced complexity mean-field model}\label{subsect:MF}

This approach is based on the introduction of the so-called quasi-invariant interaction limit \cite{Tos} which is reminiscent of the grazing collision limit for the Boltzmann equation in gas dynamics \cite{PT13,Vil}. To this end, given some small parameter $\eps \ll 1$, we consider the following rescalings of the coefficients that characterize the transitions \eqref{eq:bin_int1}--\eqref{eq:bin_int2}:
\begin{equation*}
    \lambda \mapsto \eps \lambda, \qquad \lambda_B \mapsto \eps\lambda_B, \qquad \sigma_J^2 \mapsto \eps \sigma_J^2, \qquad \sigma^2_B \mapsto \eps \sigma_B^2.
\end{equation*}
In the limit $\eps \to 0$, the microscopic interactions become quasi-invariant in the sense that we look for very small exchanges of opinion during each encounter between agents or with the background. This entails the possibility to formally derive from the Boltzmann system \eqref{eq:kinetic} a reduced complexity Fokker--Planck-type model, for each $J \in \mathcal{C}$, of the form \cite{BTZ}
\begin{equation}\label{eq:Fokker-Planck}
\begin{split}
\frac{\partial}{\partial t} f_J(w,t) &= \frac{1}{\tau^2} \bar{Q}_J(f_J,f)(w,t) + \frac{1}{\tau} \bar{Q}^B(f_J,\rho_I)(w,t) \\[4mm]
&= \frac{1}{\tau^2}\left[ \frac{\partial}{\partial w} \big(\mathcal{B}[f](w,t)f_J(w,t)\big) + \frac{\sigma_J^2}{2} \frac{\partial^2}{\partial w^2} \left(D^2(w)f_J(w,t)\right) \right] \\[4mm]
 & \qquad +\frac{1}{\tau}\left[ \lambda_B \frac{\partial}{\partial w}  \big((w - \mu(\rho_I))f_J(w,t)\big) + \frac{\sigma_B^2}{2} \frac{\partial^2}{\partial w^2}  \left(D^2(w)f_J(w,t)\right) \right], \qquad t \in \R_+,\; w \in \mathcal{I},
 \end{split}
\end{equation}
where we have defined $f = \sum_{K \in \mathcal{C}} f_K$ and the nonlocal drift operator
 \begin{equation} \label{eq:nonlocal drift}
    \mathcal{B}[f](w,t) = \lambda \sum_{K \in \mathcal{C}}  \int_{\mathcal{I}} P(w,w_*) (w-w_*) f_K(w_*,t) \dd w_*, 
 \end{equation}
and we notice that the diffusion operators are weighted by the degenerate opinion-dependent function $D(w)$, which modulates the self-thinking processes at the microscopic level. For any $J \in \mathcal{C}$, the Fokker--Planck operators \eqref{eq:Fokker-Planck} are then complemented with the no-flux boundary conditions
\begin{equation} \label{eq:Fokker-Planck BC}
\begin{split}
\mathcal{B}[f](w,t)f_J(w,t) + \frac{\sigma_J^2}{2}\dfrac{\partial}{\partial w} \left(D^2(w)f_J(w,t)\right) \Bigg|_{w= \pm1} = 0, & \\[2mm]
\lambda_B(w - \mu(\rho_I))f_J(w,t) + \frac{\sigma_B^2}{2}\dfrac{\partial}{\partial w} \left(D^2(w)f_J(w,t)\right) \Bigg|_{w =\pm 1}=0, & \\[4mm]
D^2(w) f_J(w,t) \Big|_{w = \pm 1} = 0, & 
\end{split}
\end{equation}
holding for any $t \in \R_+$.

The advantage of studying a simplified setting where the evolution of the multi-agent system is governed by the Fokker--Planck equations \eqref{eq:Fokker-Planck}--\eqref{eq:Fokker-Planck BC} is the possibility to explicitly characterize the structure of the kinetic equilibria. Specifically, we call local equilibrium of the multi-agent system any vector distribution function $(f_J^\eq)_{J \in \mathcal{C}}$ whose components $f_J^\eq = f_J^\eq(w,t)$ satisfy the normalization conditions $\int_{\mathcal{I}} f_J^\eq(w,t) \dd w = \rho_J(t)$ together with the relations
\begin{equation*}
    \mathcal{B}[f^\eq](w,t) f_J^\eq(w,t) + \frac{\sigma_J^2}{2}\frac{\partial}{\partial w}\left(D^2(w)f_J^\eq(w,t)\right) = 0, \qquad J \in \mathcal{C},
\end{equation*}
where $f^\eq = \sum_{K \in \mathcal{C}} f_K^\eq$, which shows that the components of the equilibrium state are all coupled through the nonlocal operator \eqref{eq:nonlocal drift}. Therefore, recalling the conservation law $\rho(t) = 1$ for any $t \in \R_+$ satisfied by system \eqref{eq:Fokker-Planck}, we may decouple these equilibrium relations by considering $P \equiv 1$, which allows to reduce the nonlocal drift coefficient to the form
\begin{equation*}
    \mathcal{B}[f^\eq](w,t) = \lambda (w-m(t)),
\end{equation*}
corresponding to a pure alignment-like term around the total mean opinion $m(t)$. Under this hypothesis, the equilibrium distributions can be explicitly determined by computing the indefinite integral
\begin{equation*}
f_J^\eq(w,t) = C_J(t) \exp\left(-\frac{2\lambda}{\sigma_J^2} \int \frac{w - m(t) + \frac{\sigma_J^2}{2 \lambda} \frac{\partial}{\partial w} D^2(w)}{D^2(w)} \dd w \right),
\end{equation*}
where $C_J(t) > 0$ is a suitable normalization function ensuring that $\int_{\mathcal{I}} f_J^\eq(w,t) \dd w = \rho_J(t)$ for any $J \in \mathcal{C}$. Recalling that we chose $D^2(w) = 1-w^2$, we deduce that the local equilibria correspond to beta distributions
\begin{equation} \label{eq:local equilibria}
f_J^\eq(w,t) = C_J(t) (1-w)^{-1+\frac{1-m(t)}{\nu_J}} (1+w)^{-1+\frac{1+m(t)}{\nu_J}}, \qquad J \in \mathcal{C},
\end{equation}
centered in the total mean opinion $m(t)$, where we have defined $\nu_J = \frac{\sigma_J^2}{\lambda}$ and the normalization coefficients read
\begin{equation*}
    C_J(t) = \rho_J(t) \frac{2^{-1+\frac{2}{\nu_J}}}{B\left(\frac{1-m(t)}{\nu_J},\frac{1+m(t)}{\nu_J}\right)},
\end{equation*}
with $B$ denoting the beta function, see \cite{Tos,Zan}. In the following, we will exploit these equilibria to derive new opinion-epidemiological hydrodynamic equations that will be useful to perform our concluding investigations on real data.

\subsection{The full opinion-epidemiological model}

We are now ready to introduce that complete model coupling opinion evolution with epidemic dynamics. As previously discussed, we will consider a simple $SIR$ compartmentalization of the society, but we highlight that our approach could be extended to account for more complex subdivisions of the population. Following the approach presented in \cite{A_etal,BTZ,PSHZ,Zan} we obtain the system of equations
\begin{equation}\label{eq:kinetic model}
\begin{split}
    \frac{\partial}{\partial t} f_S(w,t) &= -K(f_S,f_I)(w,t) + \frac{1}{\tau^2} \bar{Q}_S(f_S,f)(w,t) + \frac{1}{\tau} \bar{Q}_S^B(f_S,\rho_I)(w,t), \\[4mm]
    \frac{\partial}{\partial t} f_I(w,t) &= K(f_S,f_I)(w,t) - \gamma f_I(w,t) + \frac{1}{\tau^2} \bar{Q}_I(f_I,f)(w,t) + \frac{1}{\tau} \bar{Q}_I^B(f_I,\rho_I)(w,t) \\[4mm]
    \frac{\partial}{\partial t} f_R(w,t) &= \gamma f_I(w,t) + \frac{1}{\tau^2} \bar{Q}_R(f_R,f)(w,t) + \frac{1}{\tau} \bar{Q}^B(f_R,\rho_I)(w,t), \qquad t \in \R_+,\; w \in \mathcal{I},
\end{split}
\end{equation}
where we recall that we have defined $f = \sum_{K \in \mathcal{C}} f_K$. Here, the parameter $\gamma > 0$ quantifies the mean infectious period of the disease \cite{DH}, and the operator 
\begin{equation} \label{eq:epidemiological operator}
    \begin{split}
        K(f_S,f_I)(w,t) &= f_S(w,t) \int_{\mathcal{I}} \kappa(w,w_*) f_I(w_*,t) \dd w_*, 
        \end{split}
\end{equation}
defines the local incidence rate governing infection transmission. Here, $\kappa(w,w_*)$ is a nonnegative decreasing function measuring the impact of protective behaviors on the interactions between susceptible and infectious agents. A leading example for this function is given by \cite{Zan} 
\begin{equation} \label{eq:beta_T}
    \kappa(w,w_*) = \beta (1-w)^\alpha (1-w_*)^\alpha,
\end{equation}
where the constants $\beta > 0$ and $\alpha > 0$ denote respectively the baseline transmission rate of the epidemic and the exponent characterizing the efficacy of the protective measures. Thus, the local incidence rate of the epidemic (and consequently the evolution of the disease itself) is fully dependent on the agents' collective opinion regarding the use of protective behaviors. In particular, in \eqref{eq:kinetic model} the opinion dynamics are governed by the Fokker--Planck operators $\bar{Q}_J(f_J,f)$ and $\bar{Q}^B(f_J,\rho_I)$ defined by \eqref{eq:Fokker-Planck}--\eqref{eq:nonlocal drift}, which come from a quasi-invariant approximation of the agent-based interactions \eqref{eq:bin_int1}--\eqref{eq:bin_int2}.

\begin{table}
\centering
\caption{Main mathematical quantities and their public-health interpretation.}
\label{tab:public-health-interpretation}
\renewcommand{\arraystretch}{1.25}
\begin{tabular}{p{0.16\textwidth} p{0.34\textwidth} p{0.42\textwidth}}
\hline
\textbf{Notation} & \textbf{Mathematical meaning} & \textbf{Public-health interpretation} \\
\hline
$J\in\{S,I,R\}$ 
& Epidemiological compartment 
& Health status of an individual: susceptible, infectious, or removed from the transmission chain. \\

$\rho_J(t)$ 
& Fraction of the population belonging to compartment $J$ 
& Proportion of susceptible, infectious, or removed individuals at time $t$. \\

$w\in[-1,1]$ 
& Continuous individual opinion variable 
& Individual propensity to adopt protective behaviors. Values near $-1$ indicate low propensity, whereas values near $1$ indicate high propensity. \\

$f_J(w,t)$ 
& Opinion distribution of agents in compartment $J$ 
& Distribution of protective-behavior propensities among individuals with epidemiological status $J$. \\

$m_J(t)$ 
& Mean opinion in compartment $J$ 
& Average propensity to adopt protective behaviors among individuals in compartment $J$. \\

$m(t)$ 
& Mean opinion in the whole population 
& Population-average propensity to adopt protective behaviors. \\

$\mu(\rho_I(t))$ 
& Mean opinion induced by the epidemic-dependent background 
& Behavioral response induced by the observed disease burden, including risk awareness and public-health communications. \\

$\beta$ 
& Baseline transmission parameter 
& Transmission intensity in the absence of behavioral modulation. \\


$\gamma$ 
& Recovery rate
& Rate at which infectious individuals leave the infectious compartment; $1/\gamma$ is the mean infectious period. \\

\hline
\end{tabular}
\end{table}

\section{From individual-based dynamics to observable patterns}\label{sec:macro}

While the agent-based kinetic system \eqref{eq:kinetic model} provides a highly detailed mathematical description, it may be impractical for applications to real-life scenarios, notably when one wishes to calibrate the different parameters appearing in the system. In fact, it is usually important to reduce the complexity of a model, in view of lowering the computational cost of its simulation. A common way of achieving this goal in kinetic theory is to resort to the moment-based macroscopic equations associated with the mesoscopic model \eqref{eq:kinetic model}. The purpose of these kinds of derivations is precisely to characterize evolution equations for some relevant, macroscopically observable quantities (like the number of individuals or their average opinion in each compartment) that may be used for applications in place of the original kinetic system \cite{DPTZ21}. However, in a general setting, such equations cannot be determined explicitly and are not closed, in the sense that one needs to compute higher order moments to solve the equations governing the lower order ones. We therefore consider a particular case where it is possible to solve the moment equations by resorting to an equilibrium closure \cite{DPTZ21,DTZ}. More precisely, we consider the full kinetic system \eqref{eq:kinetic model} in which we fix for simplicity
\begin{equation*}
    \kappa(w,w_*) = \beta (1-w) (1-w_*), \qquad w,w_* \in \mathcal{I},
\end{equation*}
an extension has been provided in \cite{MZ23}. Furthermore, to obtain analytically the large time behavior of the mean-field approximation in Section \ref{subsect:MF} we will fix a constant interaction function $P \equiv 1$. With this choice, the nonlocal drift term \eqref{eq:nonlocal drift} reduces to
\begin{equation} \label{eq:local drift}
    \mathcal{B}[f](w,t) = \lambda (w - m(t)).
\end{equation}

In Table \ref{tab:public-health-interpretation} we recall the main mathematical quantities of the proposed modelling approach together with a brief public-health interpretation.

\subsection{Far-from-equilibrium macroscopic dynamics}

We begin by investigating the evolution of the macroscopic observables in a kinetic regime that is far from equilibrium. We initially observe that for any $J \in \mathcal{C}$ the Fokker--Planck operators $\bar{Q}_J(f_J,f)$ and $\bar{Q}^B(f_J,\rho_I)$ introduced in \eqref{eq:Fokker-Planck} are mass preserving thanks to the no-flux boundary conditions \eqref{eq:Fokker-Planck BC}, meaning that
\begin{equation*}
    \int_{\mathcal{I}} \left( \frac{1}{\tau^2} \bar{Q}_J(f_J,f)(w,t) + \frac{1}{\tau} \bar{Q}^B(f_J,\rho_I)(w,t) \right) \dd w = 0,
\end{equation*}
for any $t \in \R_+$, by using an integration by parts. Therefore, integrating over $w \in \mathcal{I}$ the kinetic model \eqref{eq:kinetic model}, we obtain the following system governing the evolution of the compartmental mass fractions:
\begin{equation} \label{eq:ffe macro 0}
    \begin{split}
        &\frac{\dd}{\dd t} \rho_S = -\beta (1- m_S) (1 - m_I) \rho_S \rho_I, \\[2mm]
        &\frac{\dd}{\dd t} \rho_I = \beta (1- m_S) (1 - m_I) \rho_S \rho_I - \gamma \rho_I, \\[2mm]
        &\frac{\dd}{\dd t} \rho_R = \gamma \rho_I, \qquad t \in \R_+.
    \end{split}
\end{equation}
In particular, by summing these equations we find that the total mass of the population $\rho(t)$ defined by \eqref{eq:mass fractions} is preserved over time. We can thus fix $\rho(t) = 1$ for any $t \in \R_+$, without loss of generality. Unlike the classical SIR model, system \eqref{eq:ffe macro 0} is not closed and one needs to compute evolution equations for the missing unknowns $m_J(t)$, $J \in \mathcal{C}$. 

Now, the operators $\bar{Q}_J(f_J,f)$ and $\bar{Q}^B(f_J,\rho_I)$ are not also mean preserving, since we can easily compute using two successive integration by parts that
\begin{equation*}
    \begin{split}
        \frac{1}{\tau^2}\int_{\mathcal{I}} w \bar{Q}_J(f_J,f)(w,t) \dd w &= -\frac{\lambda}{\tau^2} \rho_J(t) (m_J(t) - m(t)), \\[4mm]
        \frac{1}{\tau}\int_{\mathcal{I}} w\bar{Q}^B_J(f_J,\rho_I)(w,t) \dd w &= -\dfrac{\lambda_B}{\tau} \rho_J(t) \big(m_J(t) - \mu(\rho_I(t))\big). 
    \end{split}
\end{equation*}
Multiplying by $w$ and integrating over $\mathcal{I}$ both sides of \eqref{eq:kinetic model}, we deduce that the evolution of the compartmental means is described by the macroscopic system
\begin{equation} \label{eq:ffe macro 1}
    \begin{split}
        &\frac{\dd}{\dd t} \rho_S m_S = -\beta (m_S - m_{S,2}) (1 - m_I) \rho_S \rho_I - \frac{\lambda_B}{\tau} \rho_S (m_S - \mu(\rho_I)) - \frac{\lambda}{\tau^2} \rho_S (m_S - m), \\[4mm]
        &\frac{\dd}{\dd t} \rho_I m_I = \beta (m_S - m_{S,2}) (1 - m_I) \rho_S \rho_I - \gamma_I \rho_I m_I  - \frac{\lambda_B}{\tau} \rho_I (m_I - \mu(\rho_I)) - \frac{\lambda}{\tau^2} \rho_I (m_I - m), \\[4mm]
        &\frac{\dd}{\dd t} \rho_R m_R = \gamma_I \rho_I m_I - \frac{\lambda_B}{\tau} \rho_R (m_R - \mu(\rho_I)) - \frac{\lambda}{\tau^2} \rho_R (m_R - m), \qquad t \in \R_+,
    \end{split}
\end{equation}
for any $\tau > 0$, where we recall that the quantities $m_{J,2}$, $J \in \mathcal{C}$, denote the second-order moments of the compartmental distributions $f_J$, given by \eqref{eq:successive moments} with $\ell = 2$. Notice that since the drift term \eqref{eq:local drift} is centered in the total mean opinion $m(t)$ of the population, the Fokker--Planck operators $\bar{Q}_J(f_J,f)$ preserve its evolution. This is not the case however for the operator $\bar{Q}^B$, which is instead expected to play a crucial role in steering the mean population opinion, based on the individuals' sentiment about the current risk of the epidemic. In particular, thanks to the conservation properties of $K(f_S,f_I)$ in \eqref{eq:epidemiological operator} and of $\bar{Q}_J(f_J,f)$, summing equations \eqref{eq:ffe macro 1} we end up with the same evolution equation \eqref{eq:evolution of m} for $m(t)$ that we determined in the Boltzmann setting.

This is however not enough to close equations \eqref{eq:ffe macro 0}, hence we are once again facing the previous issue where, in the attempt of closing the above underdetermined system \eqref{eq:ffe macro 1} it is necessary to resort to the equations governing the second-order moments $m_{J,2}(t)$, and the procedure goes on creating a so-called hierarchy of moment equations. Let us outline a possible strategy to stop this infinite loop, by looking at the evolution of the kinetic system \eqref{eq:kinetic model} near equilibrium.

\subsection{Close-to-equilibrium macroscopic dynamics}
Following the classical kinetic theory of gases, in order to close this iterative process one can employ a relaxation argument by supposing that the scaling parameter $\tau$ in \eqref{eq:kinetic model} is taken small enough, so that $f_J(w,t) \underset{\tau \ll 1}{\sim} f_J^\eq(w,t)$ for any $J \in \mathcal{C}$, where the local equilibria $f_J^\eq$ are the unique solutions with mass $\rho_J(t)$ of the relations $\bar{Q}_J(f_J^\eq,f^\eq)(w,t) = 0$, $J \in \mathcal{C}$, which are given by the beta distributions \eqref{eq:local equilibria}. In this regime $\tau \ll 1$, one can formally consider the close-to-equilibrium evolution of the kinetic system \eqref{eq:kinetic model}, given by
\begin{equation} \label{eq:cte kinetic model}
\begin{split}
    \frac{\partial}{\partial t} f_S^\eq(w,t) &= -K(f_S^\eq,f_I^\eq)(w,t) + \frac{1}{\tau} \bar{Q}^B(f_S^\eq,\rho_I)(w,t), \\[4mm]
    \frac{\partial}{\partial t} f_I^\eq(w,t) &= K(f_S^\eq,f_I^\eq)(w,t) -\gamma f_I^\eq(w,t)+ \frac{1}{\tau} \bar{Q}^B(f_I^\eq,\rho_I)(w,t) \\[4mm]
    \frac{\partial}{\partial t} f_R^\eq(w,t) &= \gamma f_I^\eq(w,t) + \frac{1}{\tau} \bar{Q}^B(f_R^\eq,\rho_I)(w,t)
    \qquad t \in \R_+,\; w \in \mathcal{I}.
    \end{split}
\end{equation}
With the new information coming from the structure of the local equilibria, one can then perform the same computations used to derive equations \eqref{eq:ffe macro 0} and \eqref{eq:ffe macro 1} to recover in a formal regime $\tau \ll 1$, a new evolution system for the compartmental mass fractions
\begin{equation} \label{eq:cte macro 0}
    \begin{split}
        & \frac{\dd}{\dd t} \rho_S = -\beta (1- m)^2 \rho_S \rho_I, \\[2mm]
        & \frac{\dd}{\dd t} \rho_I = \beta (1- m)^2 \rho_S \rho_I - \gamma \rho_I, \\[2mm]
        & \frac{\dd}{\dd t} \rho_R = \gamma \rho_I, \qquad t \in \R_+,
    \end{split}
\end{equation}
coupled with the evolution equation for the total mean opinion
\begin{equation} \label{eq:cte macro 1}
    \frac{\dd}{\dd t} m = - \frac{\lambda_B}{\tau} \big(m - \mu(\rho_I)\big), \qquad t \in \R_+,
\end{equation}
providing together a now fully closed macroscopic model. 

We can then infer several information on the evolution of the coupled system \eqref{eq:cte macro 0}--\eqref{eq:cte macro 1}. Indeed, since the coefficient $\beta (1-m(t))^2$ is always nonnegative, the behavior of system \eqref{eq:cte macro 0} follows that of the classical SIR equations. First of all, we can easily check that $0 \leq \rho_J(t) \leq 1$ for any $t \in \R_+$ and $J \in \mathcal{C}$. Moreover, $\rho_S$ is non-increasing and $\rho_R$ is non-decreasing, hence they both converge toward some limits $\rho_S^\infty, \rho_R^\infty \in [0,1]$. Thanks to mass conservation, also $\rho_I$ converges and its limit $\rho_I^\infty$ must be zero since it holds that
\begin{equation*}
    \int_{\R_+} \rho_I(t) \dd t = \frac{1}{\gamma} \left(\rho_R^\infty - \rho_R^\init\right), \qquad \left| \frac{\dd}{\dd t} \rho_I \right| < +\infty.
\end{equation*}
In particular, since $\mu \in C^1[0,1]$, we deduce that $\mu(\rho_I(t)) \underset{t \to +\infty}{\longrightarrow} \mu(0)=-1$ and the relaxation speed is the same of the convergence $\rho_I(t) \underset{t \to +\infty}{\longrightarrow} 0$. Then, system \eqref{eq:cte macro 1} can be solved explicitly, depending on the evolution of $\rho_I$, as
\begin{equation*}
    m(t) = m(0) e^{-\frac{\lambda_B}{\tau} t} + \frac{\lambda_B}{\tau} \int_0^t \mu(\rho_I(s)) e^{-\frac{\lambda_B}{\tau} (t-s)} \dd s,
\end{equation*}
showing that the total mean $m(t)$ relaxes toward $\mu(\rho_I(t))$ when $\tau \to 0$ or $t \to +\infty$.

\begin{remark}
The modified SIR system \eqref{eq:cte macro 0}--\eqref{eq:cte macro 1} is characterized by a time-dependent effective reproduction number
\begin{equation*}
    \mathcal{R}_{\mathrm{eff}}(t) = \frac{\beta}{\gamma} (1-m(t))^2 \rho_S(t),
\end{equation*}
governing the evolution of the epidemic through the influence of the total mean opinion of the population $m(t)$. In turn, the temporal changes in $m(t)$ are affected by the background opinion $\mu(\rho_I(t))$ through the evolution of the disease itself. This double feedback mechanism highlights the main interest of the model: the continuous reaction $\mu(\rho_I(t))$ of the population to the current state of the epidemic modifies the public opinion $m(t)$, while the latter acts as an intrinsic control parameter that leads to spontaneous behavioral changes capable of reducing the impact of the disease. In particular, in the relaxation limit $\tau \to 0$, the model \eqref{eq:cte macro 0}--\eqref{eq:cte macro 1} formally converges to the kinetic-controlled system
\begin{equation} \label{eq:cte limit}
    \begin{split}
        & \frac{\dd}{\dd t} \rho_S = -\beta \big(1 - \mu(\rho_I) \big)^2 \rho_S \rho_I, \\[2mm]
        & \frac{\dd}{\dd t} \rho_I = \beta \big(1 - \mu(\rho_I) \big)^2 \rho_S \rho_I - \gamma \rho_I, \\[2mm]
        & \frac{\dd}{\dd t} \rho_R = \gamma \rho_I, \qquad t \in \R_+.
    \end{split}
\end{equation}
\end{remark}

\begin{remark}
In the following section we will systematically consider as background opinion $\mu(\rho_I(t))$ the polynomial function proposed in \cite{BTZ}, which has the explicit expression
\begin{equation} \label{eq:mu}
    \mu(\rho_I(t)) = 1-2(1-\rho_I(t))^r, \qquad t \in \R_+,
\end{equation}
where $r > 0$ is a constant that characterizes the intensity of the individuals' risk perception in connection with the current state of the epidemic. With this particular modeling choice, from \eqref{eq:cte limit} we get that in the limit $\tau \to 0$ the effective transmission rate of \eqref{eq:cte macro 0} is given by
\begin{equation*}
    \beta_{\mathrm{eff}}(\rho_I(t)) = 4\beta(1-\rho_I(t))^{2r}.
\end{equation*}
We can observe how, for every $r > 0$, $\beta_{\mathrm{eff}}$ is a strictly decreasing function of the mass fraction of infected individuals, namely
\begin{equation*}
    \frac{\dd}{\dd \rho_I} \beta_{\mathrm{eff}}(\rho_I) =-8 r \beta(1-\rho_I)^{2r -1}<0, \qquad 0<\rho_I<1.
\end{equation*}
Therefore, the behavioral response encoded by $m(t) \underset{\tau \ll 1}{\sim} \mu(\rho_I(t))$ as in \eqref{eq:mu} progressively reduces the transmission rate, introducing an endogenous negative feedback into the epidemic dynamics. Moreover, for every fixed $0<\rho_I<1$,
\begin{equation*}
    \frac{\partial}{\partial r} \beta_{\mathrm{eff}}(\rho_I) = 8\beta(1-\rho_I)^{2r}\log(1-\rho_I)<0,
\end{equation*}
since $\log(1-\rho_I)<0$. Hence, increasing $r$ strengthens the reduction of the effective transmission rate induced by opinion formation.
\end{remark}

\section{Numerical results} \label{sec:numerics}

The objective of this section is twofold. First, we illustrate the qualitative behavior of the proposed opinion-epidemic model through in-silico test cases that highlight the role of collective opinion in shaping the epidemic evolution. Second, we consider a calibration of the model based on real epidemiological and social media data from the first wave of the COVID-19 pandemic in Italy, showing how the coupled framework can simultaneously infer epidemiological parameters, characterize the evolution of public risk perception, and estimate the representativeness of a biased social media sample.

Throughout our investigation we consider the hydrodynamic regime $\tau \ll 1$, in which the kinetic description is accurately approximated by the macroscopic system \eqref{eq:cte macro 0}--\eqref{eq:cte macro 1} enabling clear comparison with the available data. We recall that this corresponds to a situation where opinion exchanges occur on a much faster time scale than the epidemic evolution, allowing the compartmental opinion distributions to rapidly relax toward their local equilibria. The resulting modified SIR system \eqref{eq:cte macro 0} is integrated by means of a fourth-order Runge--Kutta method, while the stiff relaxation equation \eqref{eq:cte macro 1} is treated implicitly using a backward Euler discretisation.

\begin{figure}[h!]
\centering
\includegraphics[width = 0.32\linewidth]{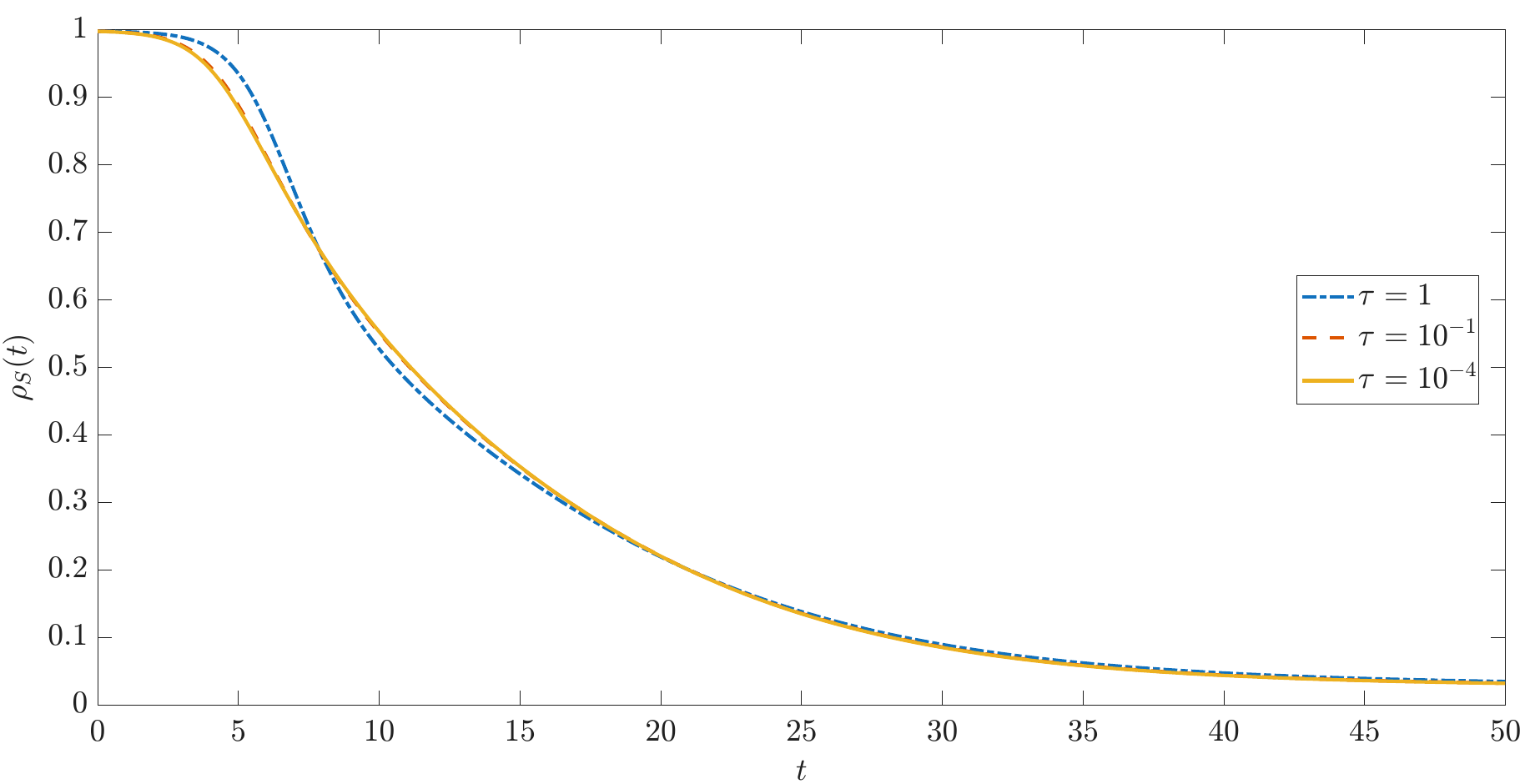} 
\includegraphics[width = 0.32\linewidth]{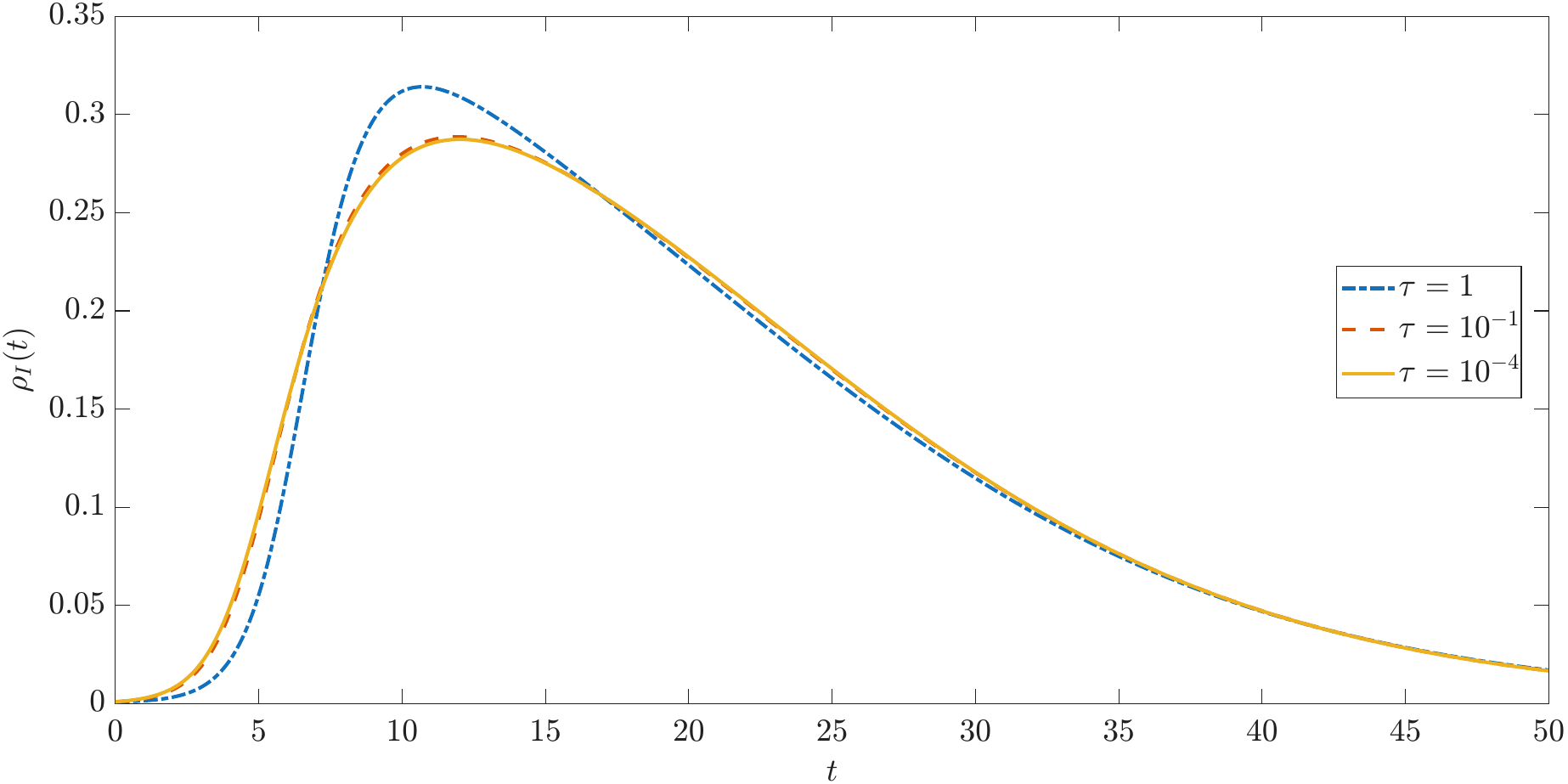} 
\includegraphics[width = 0.32\linewidth]{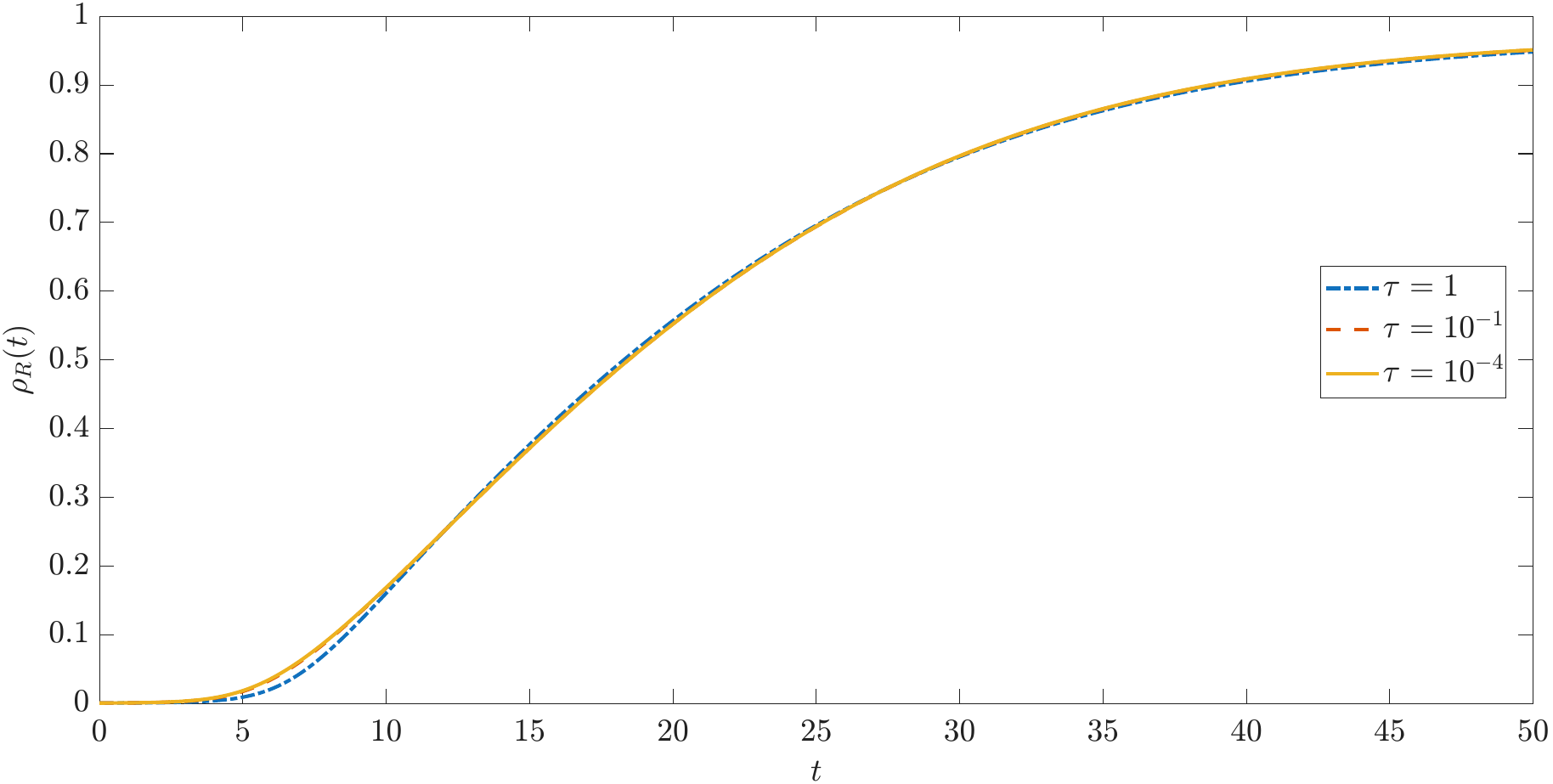} \\
\includegraphics[width = 0.32\linewidth]{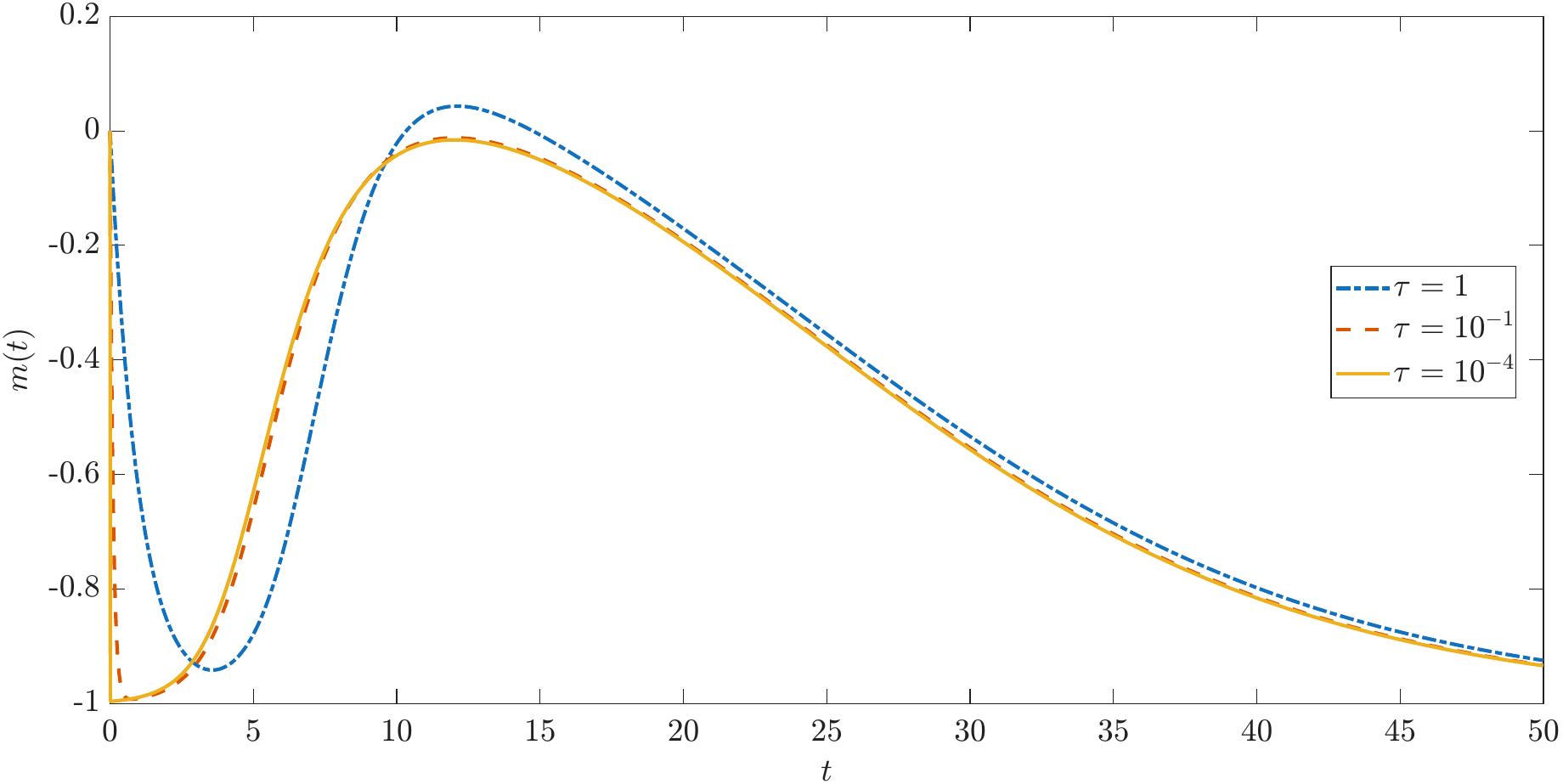} 
\includegraphics[width = 0.32\linewidth]{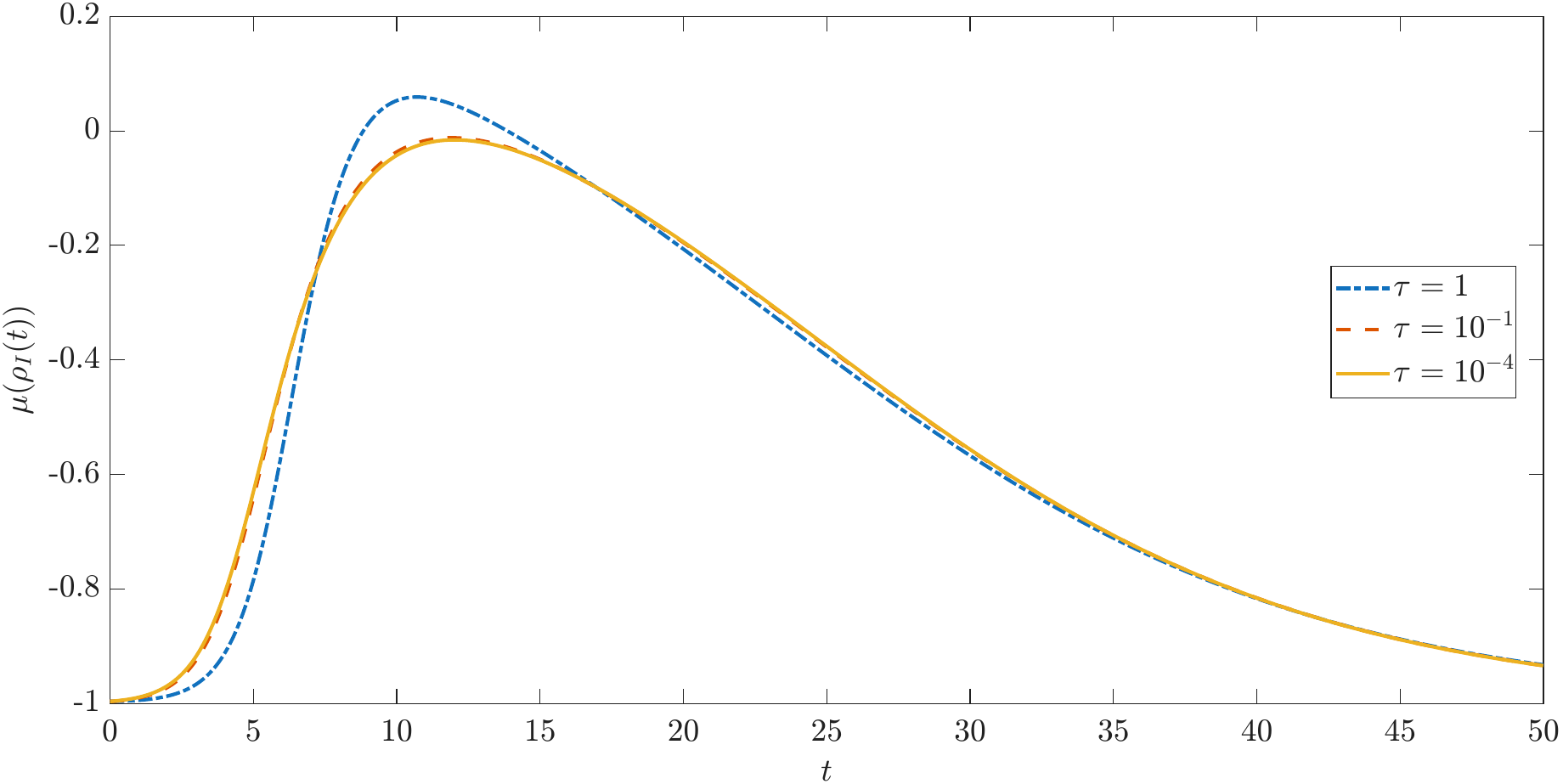}
\caption{\textbf{Test 1 -- Consistency of the relaxation limit.} Evolution of the model \eqref{eq:cte macro 0}--\eqref{eq:cte macro 1} and of the background opinion \eqref{eq:mu}, for different values of the relaxation time scale $\tau = 1, 10^{-1}, 10^{-4}$. We considered the initial conditions \eqref{eq:initial conditions}. Values of the parameters: $\beta = 0.3$, $\gamma = 1/7$, $\lambda_B = 1$, and $r = 2$ in \eqref{eq:mu}.}
\label{fig:test 1}
\end{figure}

\subsection{Impact of opinion formation dynamics on epidemic evolution}

\subsubsection{Consistency of the relaxation model}

We begin to analyze the relaxation limit $\tau \to 0$ of system \eqref{eq:cte macro 0}--\eqref{eq:cte macro 1}. For this we consider the explicit form of the background opinion $\mu(\rho_I(t))$ given by \eqref{eq:mu}, with the risk perception parameter fixed to $r = 2$. Starting from the initial conditions
\begin{equation} \label{eq:initial conditions}
\rho_I(0) = \rho_R(0) = 10^{-3},\qquad \rho_S(0) = 1 - \rho_I(0) - \rho_R(0), \qquad m(0) = 0,
\end{equation}
we let system \eqref{eq:cte macro 0}--\eqref{eq:cte macro 1} evolve for different, decreasing values of $\tau = 1, 10^{-1}, 10^{-4}$ and we plot the results in Figure \ref{fig:test 1}. We notice that when $\tau = 10^{-4}$ (continuous yellow curves) the model has already relaxed to the limit system \eqref{eq:cte limit} and the total mean opinion $m(t)$ follows exactly the evolution of the background opinion $\mu(\rho_I(t))$, apart from a small initial layer of order $\mathcal{O}(\tau^{1/2})$ accounting for the non-well-prepared initial condition $m(0) = 0$ (note in fact that $\mu(\rho_I(0)) \approx -1$).

\subsubsection{Influence of the background opinions on the epidemic trends}

We proceed to investigate the impact of the background opinion $\mu(\rho_I(t))$ on the evolution of system \eqref{eq:cte macro 0}--\eqref{eq:cte macro 1}. We perform two tests where we consider in both cases the relaxation parameter $\tau = 10^{-4}$ and the initial conditions \eqref{eq:initial conditions}.

\begin{figure}[h!]
\centering
\includegraphics[width = 0.48\linewidth]{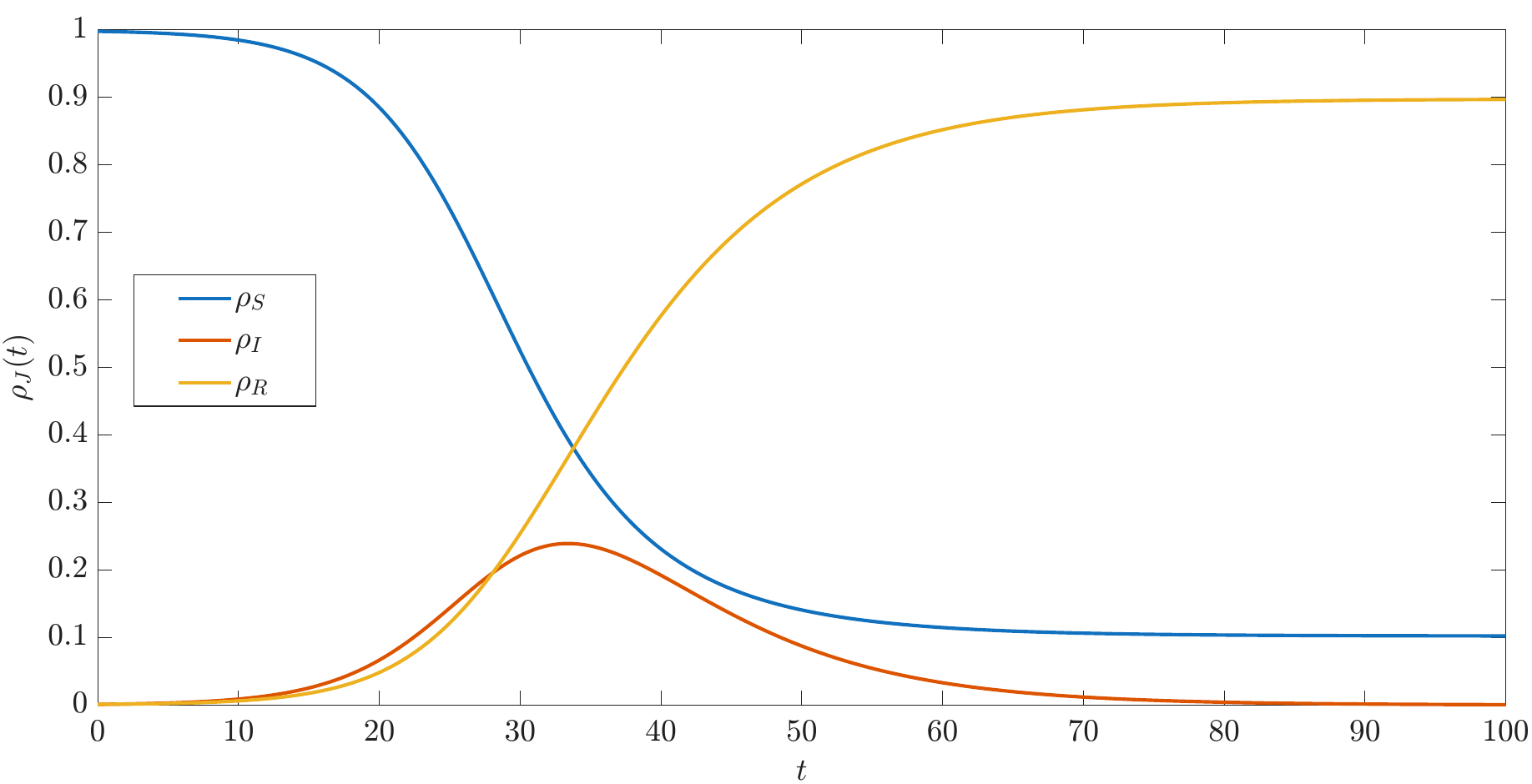} \hfill
\includegraphics[width = 0.48\linewidth]{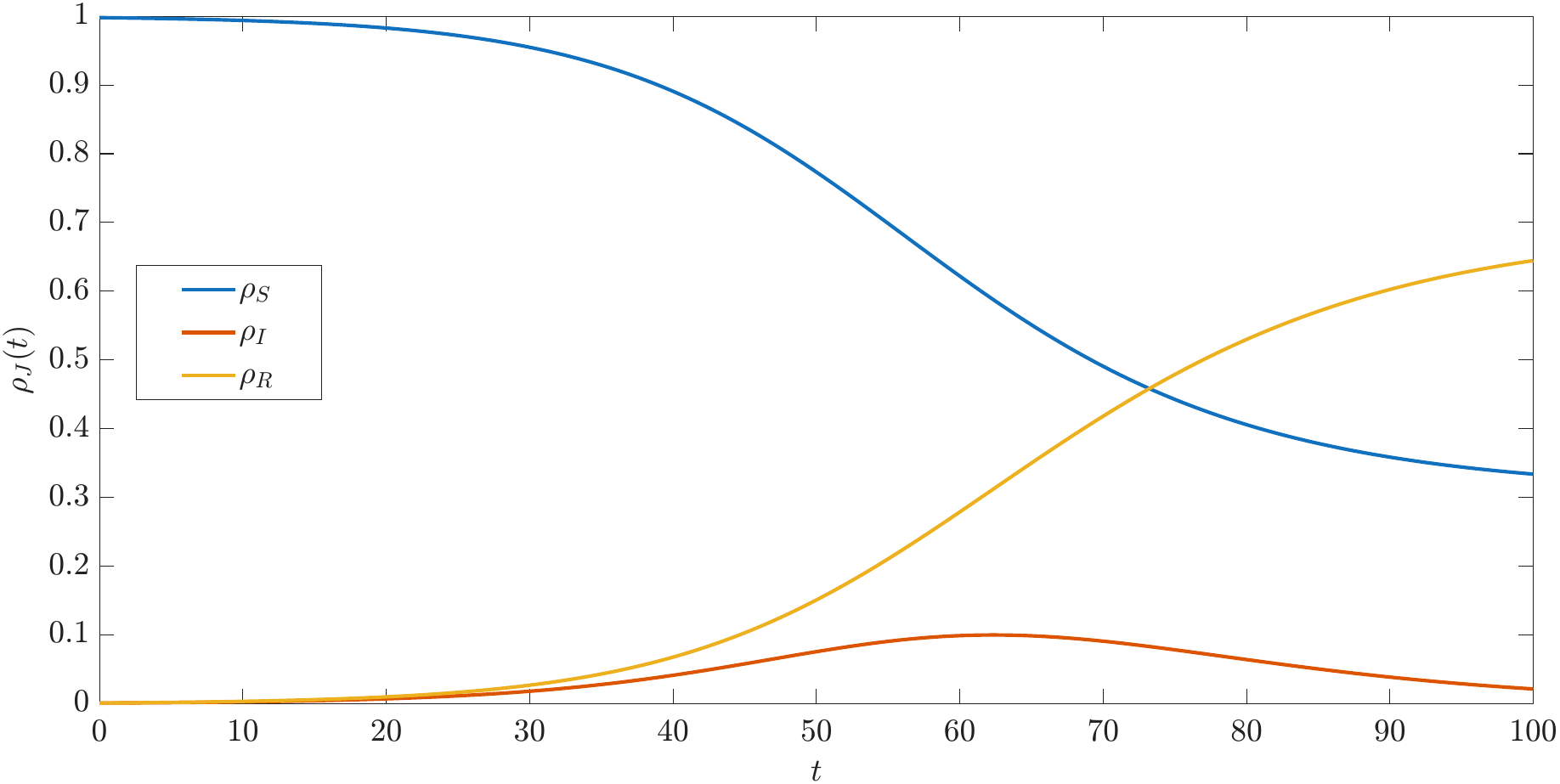}
\caption{\textbf{Test 2 -- Influence of $\bm{\mu(\rho_I(t))}$.} Evolution of the model \eqref{eq:cte macro 0}--\eqref{eq:cte macro 1} in the case of a constant negative background opinion $\mu = -0.1$ (left) or a positive background opinion $\mu = 0.1$  (right). We considered the initial conditions \eqref{eq:initial conditions}. Values of the parameters: $\beta = 0.3$, $\gamma = 1/7$, $\lambda_B = 1$, and $\tau = 10^{-4}$.}
\label{fig:test 2}
\end{figure}

In the first test we assume that $\mu$ is a constant function and we compare the evolution of the mass fractions $\rho_J(t)$ when a positive or a negative value of the background opinion is considered. The results are displayed in Figure \ref{fig:test 2}. We observe that for negative values ($\mu(\rho_I) \equiv -0.1$, left figure) the infection peak is more pronounced and occurs earlier during the epidemic, while for positive values ($\mu(\rho_I) \equiv 0.1$, right figure) the infection takes much more time to develop and its impact on the population is lessened. This is due to the fact that the mean population opinion $m(t)$ is instantaneously steered by equation \eqref{eq:cte macro 1} to the chosen constant value of $\mu$, which acts directly as a negative/positive control parameter on the evolution of the SIR system \eqref{eq:cte macro 0}.

\begin{figure}[h!]
\centering
\includegraphics[width = 0.48\linewidth]{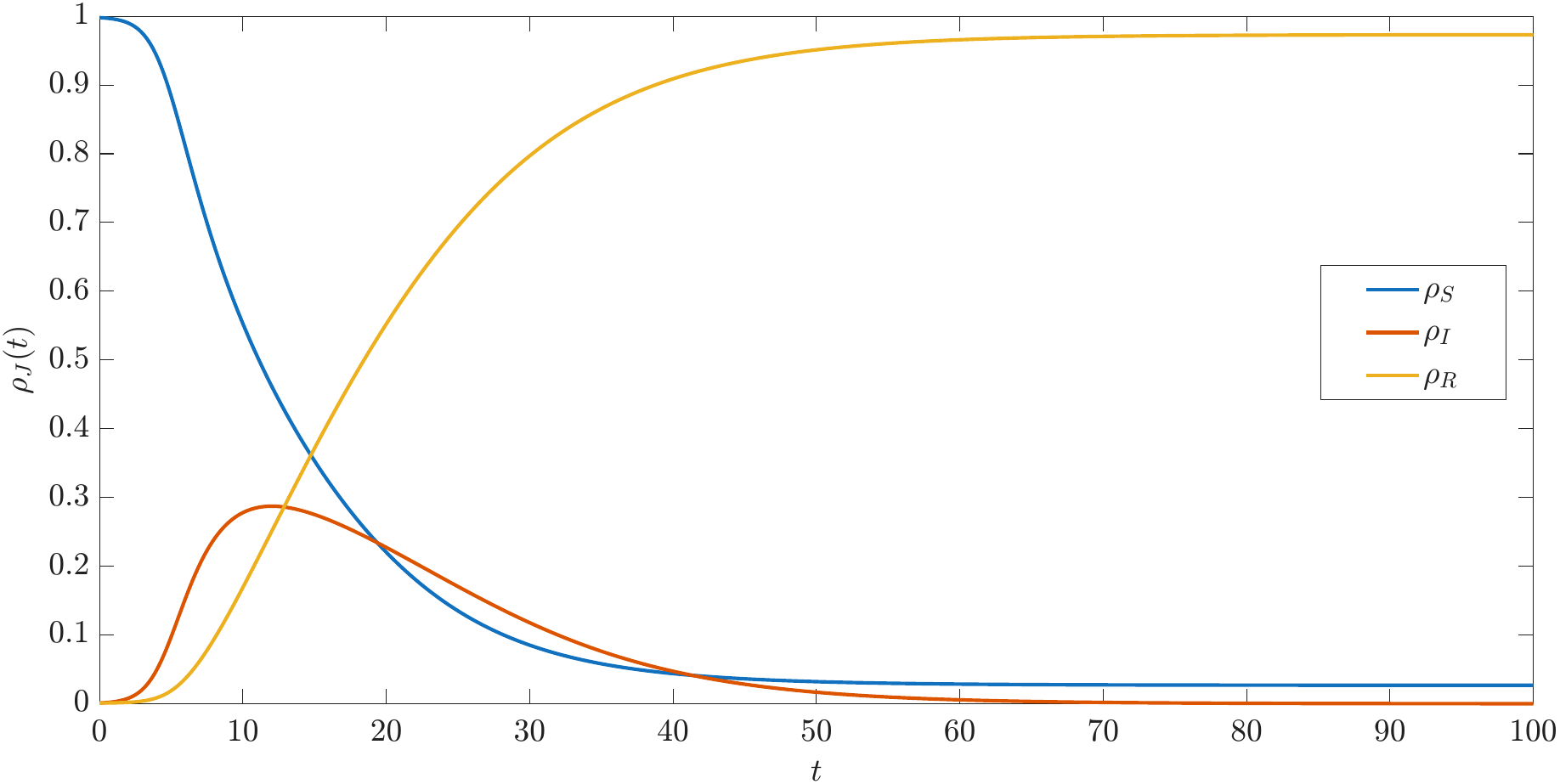} \hfill
\includegraphics[width = 0.48\linewidth]{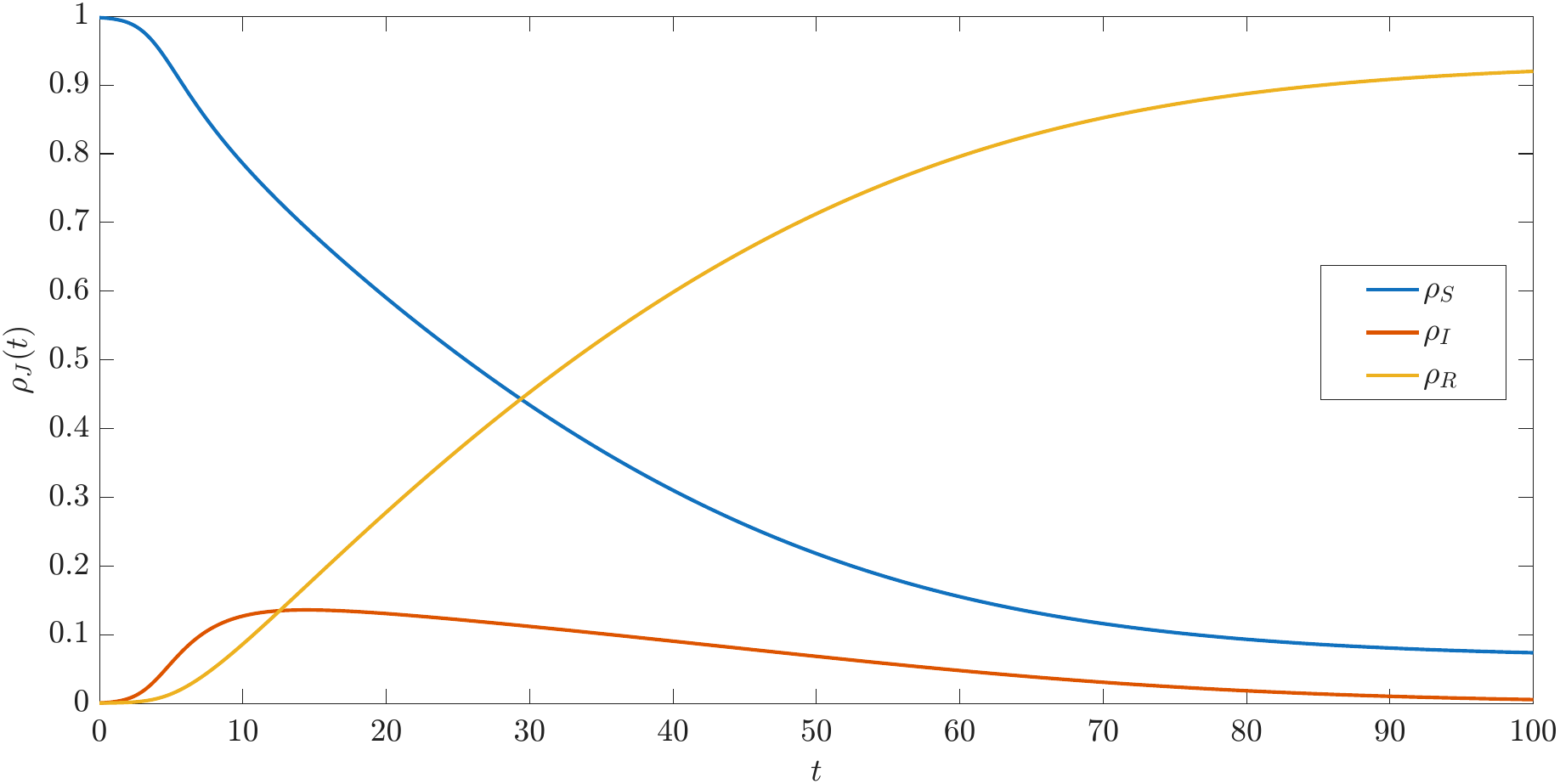}
\caption{\textbf{Test 3 -- Influence of the parameter $\mathbf{r}$.} Evolution of the model \eqref{eq:cte macro 0}--\eqref{eq:cte macro 1} when the background opinion evolves according to \eqref{eq:mu}, with weak (left) or strong reaction (right), respectively obtained setting the risk perception parameter to $r = 2$ or $r = 6$. We considered the initial conditions \eqref{eq:initial conditions}. Values of the parameters: $\beta = 0.3$, $\gamma = 1/7$, $\lambda_B = 1$, and $\tau = 10^{-4}$.}
\label{fig:test 3}
\end{figure}

In the second test, we consider the specific form of the background opinion $\mu(\rho_I(t))$ given by \eqref{eq:mu}. In particular, recall that the exponent $r$ of this polynomial function models the individuals' risk perception, namely the strength of the population's reaction to the current epidemic state $\rho_I(t)$, with greater values of $r$ leading to stronger effects of the reaction. In Figure \ref{fig:test 3}, we compare the evolution of the compartmental mass fractions for two values of this parameter. We notice that for lower values of risk perception ($r = 2$, left figure) the population reacts weakly to the epidemic spread, giving rise to a very steep initial increase in the number of infected individuals. In particular, the feedback of $\mu(\rho_I(t))$ on the evolution of the SIR system \eqref{eq:cte macro 0} takes more time to start taking effect, allowing the disease to affect a larger portion of the population. On the contrary, for bigger values of risk perception ($r = 6$, right figure) the population is able to adapt faster to the initial growth in the number of infected individuals and its stronger reaction allows for a significant decrease of the infection peak. Notice in particular that, although the disease lasts for a longer time, the overall number of recovered individuals (hence, the impact of the epidemic) is visibly reduced.

\subsection{Opinion dynamics and early phases of the COVID-19 pandemic in Italy}

We now assess the predictive capability of the proposed framework using epidemiological and social media data from the first wave of the COVID-19 epidemic in Italy. The epidemiological observations consist of the daily numbers of infected and recovered individuals between February 24 and April 25, 2020. To characterize the evolution of the public opinion we considered the Twitter dataset analyzed in \cite{Rizzo_etal} which contains data for the period December 1, 2019 and April 25, 2020. The available data consist of Twitter (now X) posts in Italian related to the COVID-19 pandemic that were classified through a machine-learning-based sentiment analysis developed and validated in \cite{Rizzo_etal}. In particular, we consider the daily average sentiment score provided by their analysis as a proxy for the opinion expressed by the corresponding social media community. Still, we acknowledge that these data do not represent the opinion of the Italian population as a whole. Rather, they capture the sentiment of a specific subset of users active on Twitter.

\begin{figure}
\centering
\includegraphics[width = \linewidth]{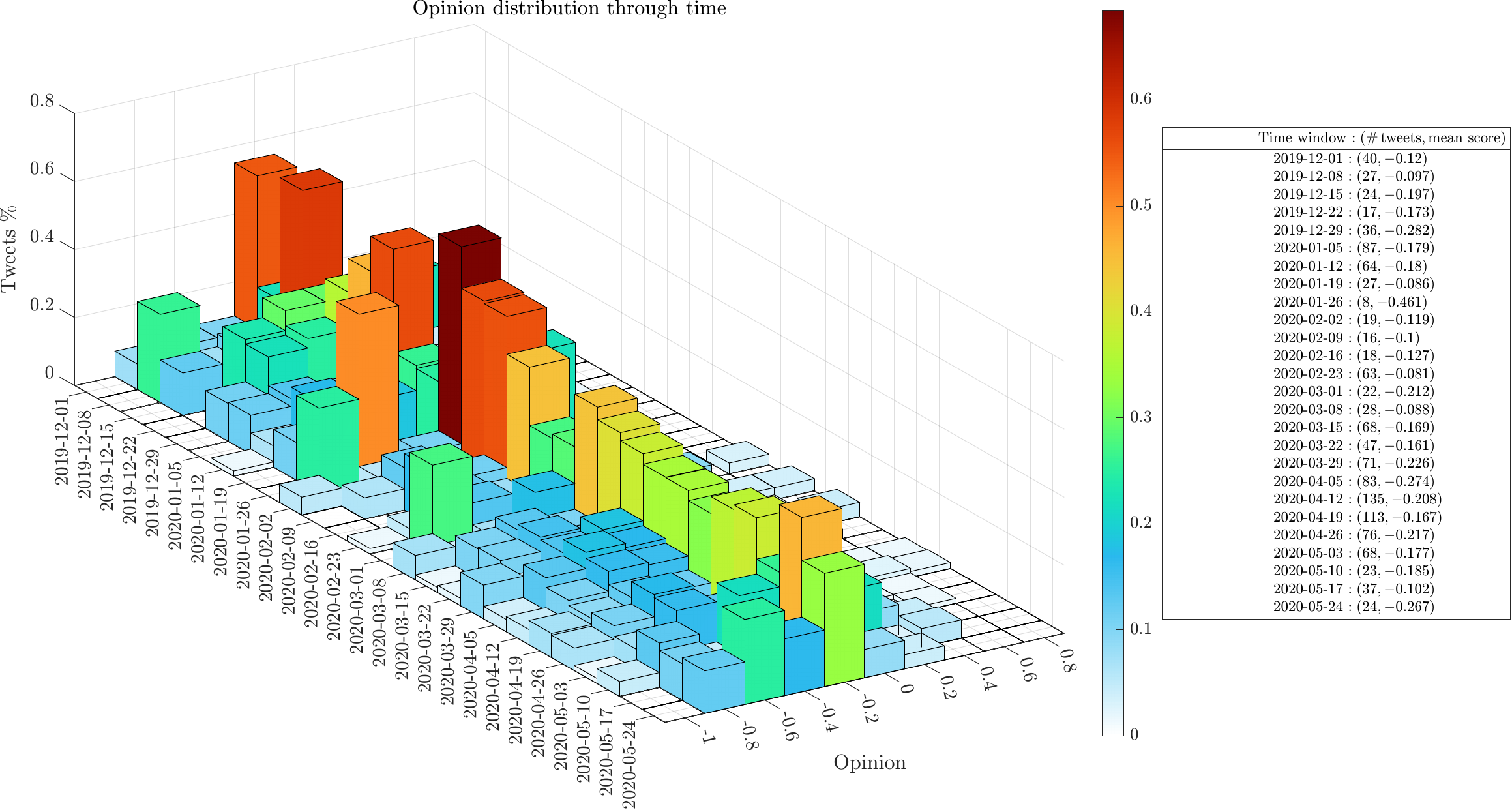}
\caption{\textbf{Empirical data.} Evolution of the empirical opinion distribution from social media data analyzed in \cite{Rizzo_etal}.}
\label{fig:dist_opinion}
\end{figure}

Based on the data provided in \cite{Rizzo_etal}, we reconstruct the empirical opinion distribution by aggregating the sentiment scores over consecutive seven-day windows. Figure \ref{fig:dist_opinion} shows the resulting evolution of the opinion distribution together with the number of posts analyzed in each week. 

\begin{figure}[h!]
\centering
\includegraphics[width = 0.48\linewidth]{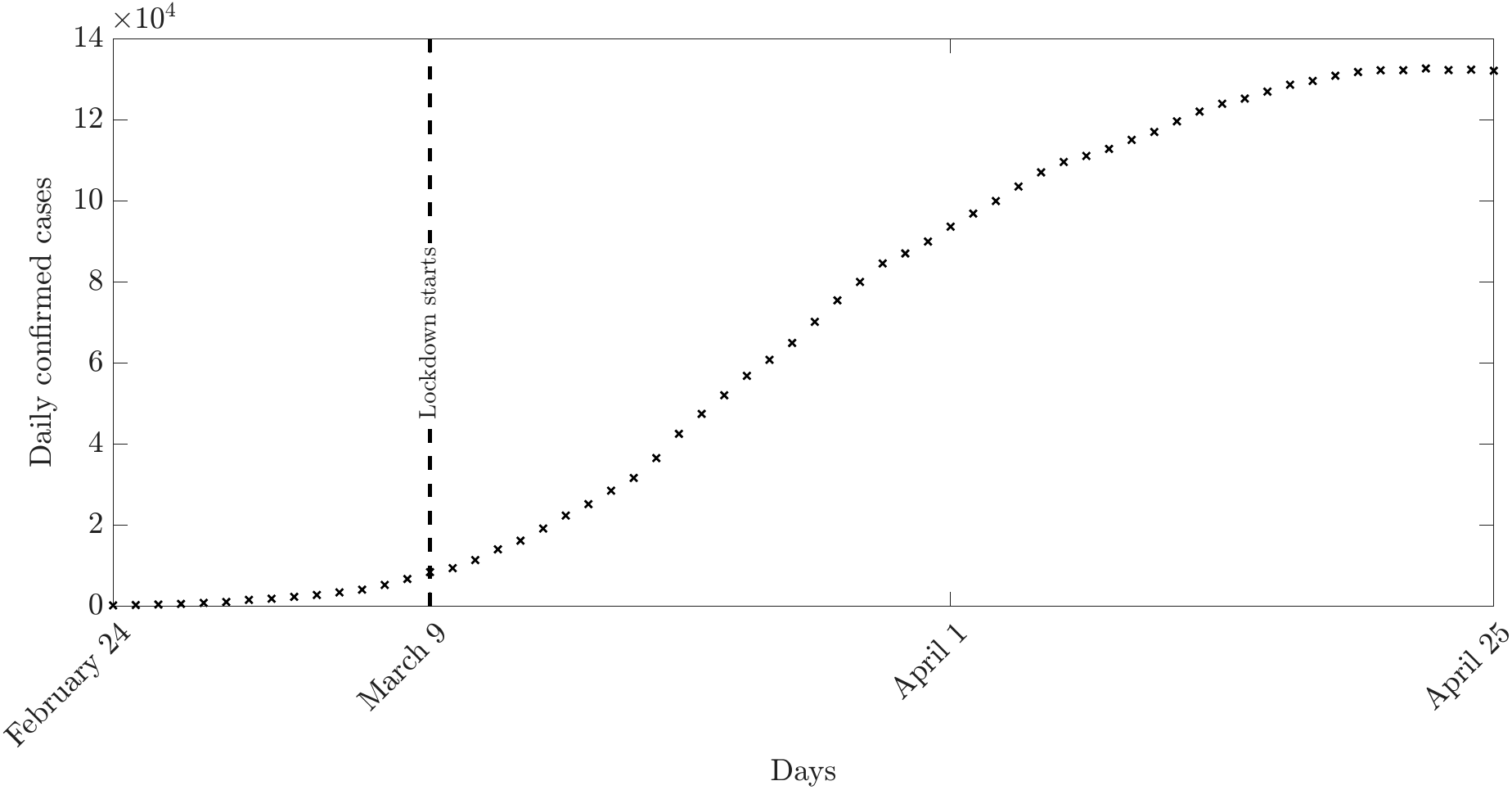} \hfill
\includegraphics[width = 0.48\linewidth]{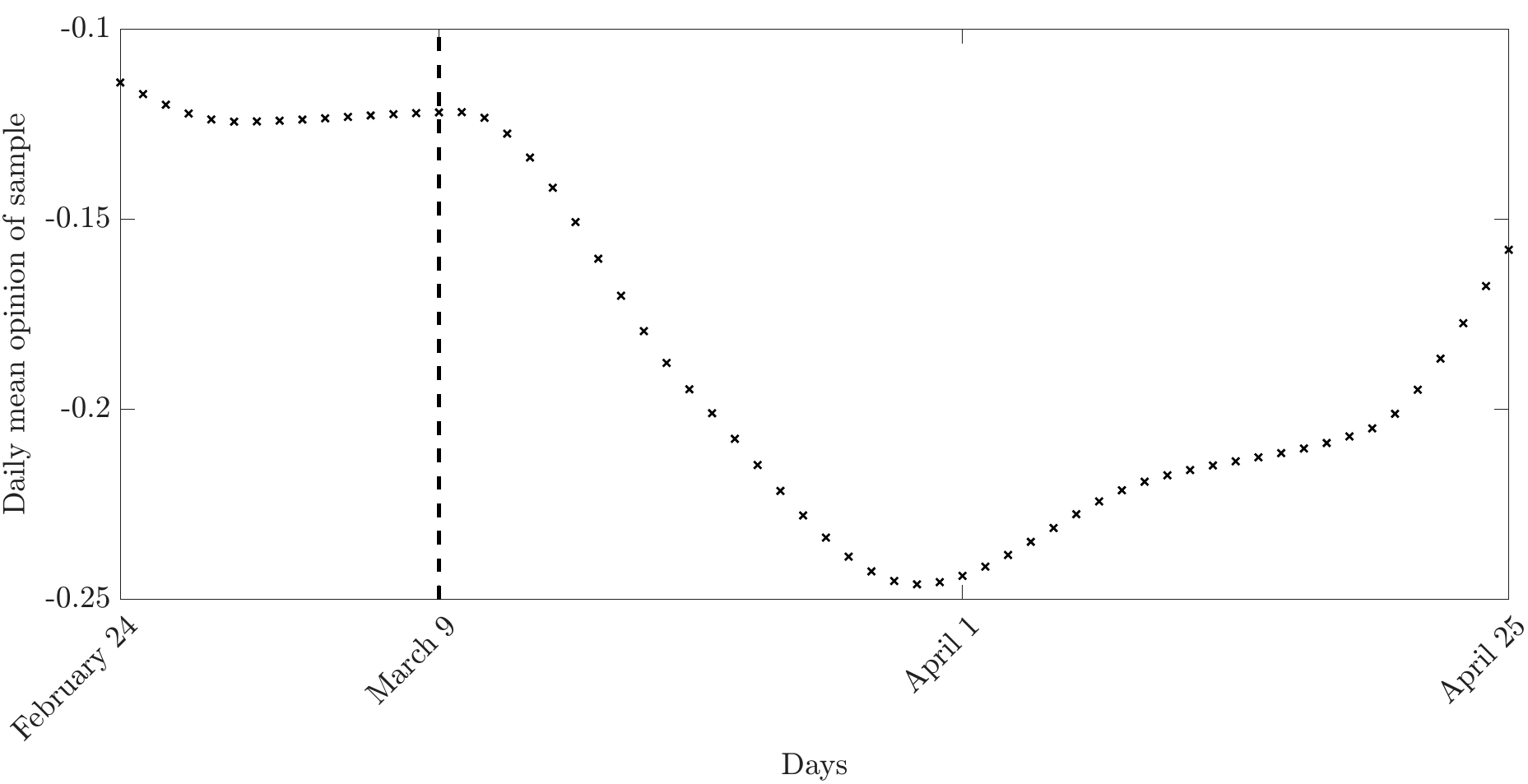}
\caption{\textbf{Empirical data.} Evolution of reported active COVID-19 cases in Italy (left) during the initial months of the pandemic and average sentiment of one part of the population toward the containment measures (right) during the same period, determined from a sample of Twitter posts on this topic.}
\label{fig:Data}
\end{figure}

From the epidemiological point of view, the relevant data we consider are the number of infected and recovered individuals over the time span February 24--April 25, 2020. To enable a daily comparison between epidemiological and opinion data, we interpolate the obtained mean opinion of the sample of Twitter posts. The  evolution of the total number of infected and the trend of the observed mean opinion are depicted in Figure \ref{fig:Data}.

\subsubsection{Estimation of the risk perception}

In the following section, we seek to obtain a data-based understanding of the functional relation between the current number of infected individuals $\rho_I(t)$ and the risk perception $\mu(\rho_I(t))$. To this end, we exploit the macroscopic model \eqref{eq:ffe macro 0}--\eqref{eq:cte macro 1} to infer the optimal values of the epidemiological parameters $\beta, \gamma > 0$ before the lockdown started on March 9, 2020 in Italy. 

\begin{figure}
\centering
\includegraphics[width = 0.48\linewidth]{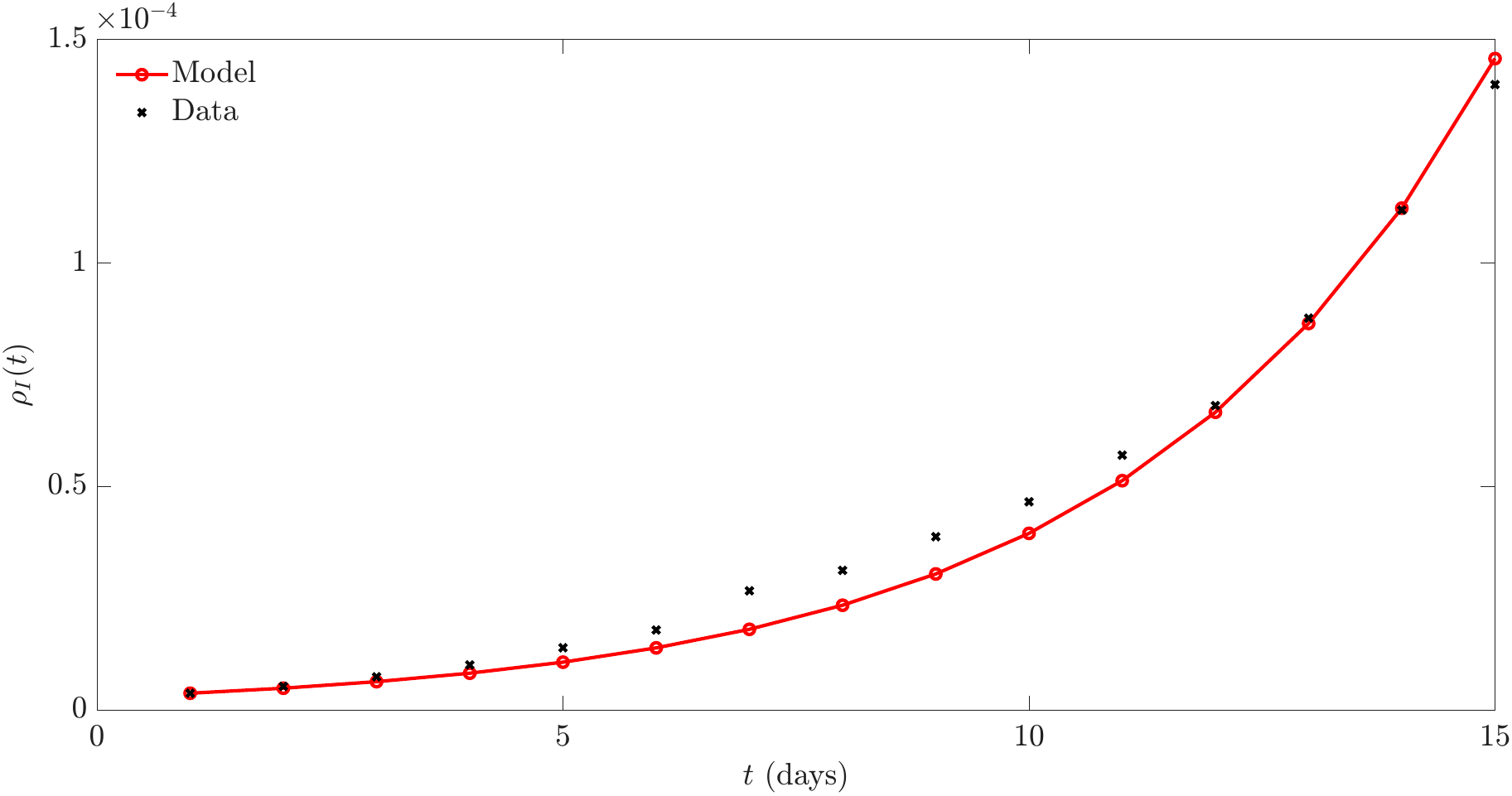} \hfill
\includegraphics[width = 0.48\linewidth]{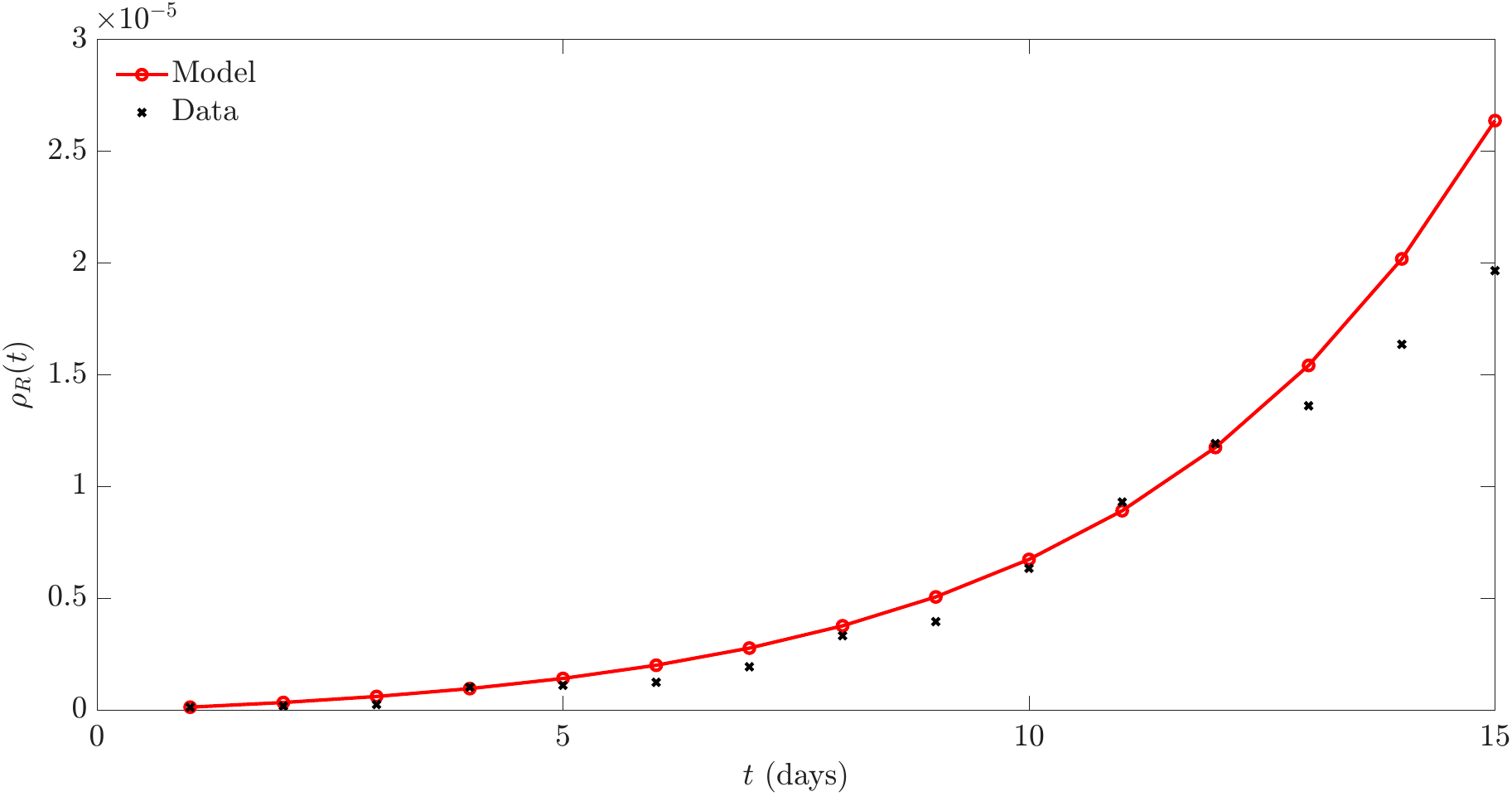}
\caption{\textbf{Pre-lockdown analysis.} Model results using the optimized parameters from Table \ref{tab:inferred-parameters}. Comparisons between the data and their approximation provided by the model, for the number of infected $\rho_I(t)$ (left) and recovered $\rho_R(t)$ (right) individuals in Italy, during the two weeks preceding the first lockdown.}
\label{fig:Predictions-BL}
\end{figure}

At this initial stage we fix the function $\mu(\rho_I) \equiv -1$, to ensure that all information on the initial spreading dynamics are correctly encapsulated by the sole parameter $\beta$. In particular, we proceed as in \cite{albi_21,ZBDDPAFT} and seek to minimize a convex combination of the $L^2$ distances between our data and their approximations computed by the model, which are the mass fractions of infected $\rho_I(t)$ and recovered $\rho_R(t)$ individuals. We collect all the obtained relevant parameters in Table \ref{tab:inferred-parameters}, and the resulting model predictions are displayed in Figure \ref{fig:Predictions-BL}.

\begin{table}[h!]
\begin{center}
\begin{tabular}{| c | c | c |}
\hline
\cellcolor{lightgray}{Parameter} & \cellcolor{lightgray}{Meaning} & \cellcolor{lightgray}{Value} \\
$\beta$ & baseline transmission rate & 0.0773 \\
$\gamma$ & average recovery rate & 0.0482 \\
$r$ & intensity of risk perception  & $ 4.23 \times 10^2 $ \\
\hline
\end{tabular}
\end{center}
\caption{Epidemiological parameters inferred from the data.}
\label{tab:inferred-parameters}
\end{table}

\begin{figure}
\centering
\includegraphics[width = 0.48\linewidth]{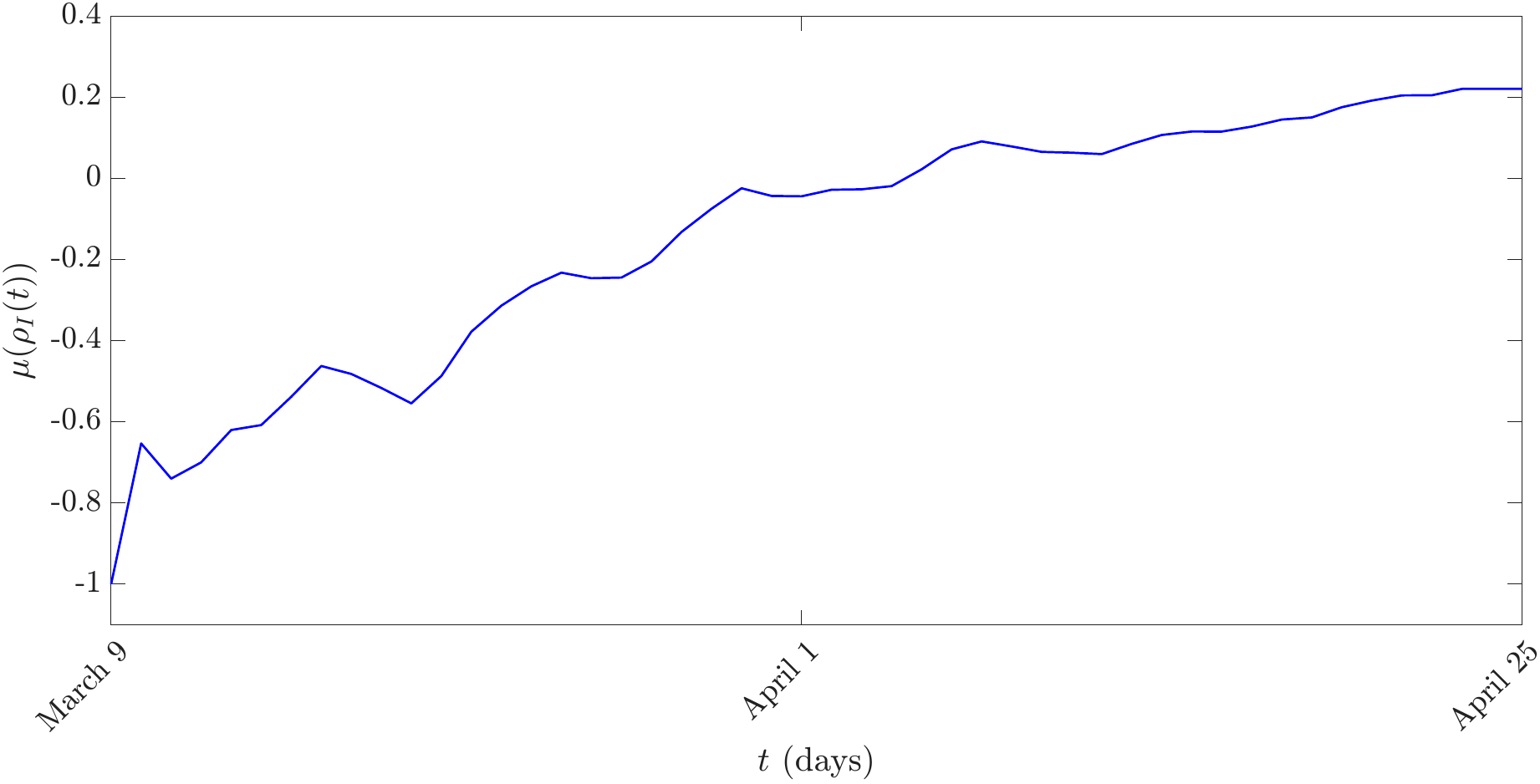} \hfill
\includegraphics[width = 0.48\linewidth]{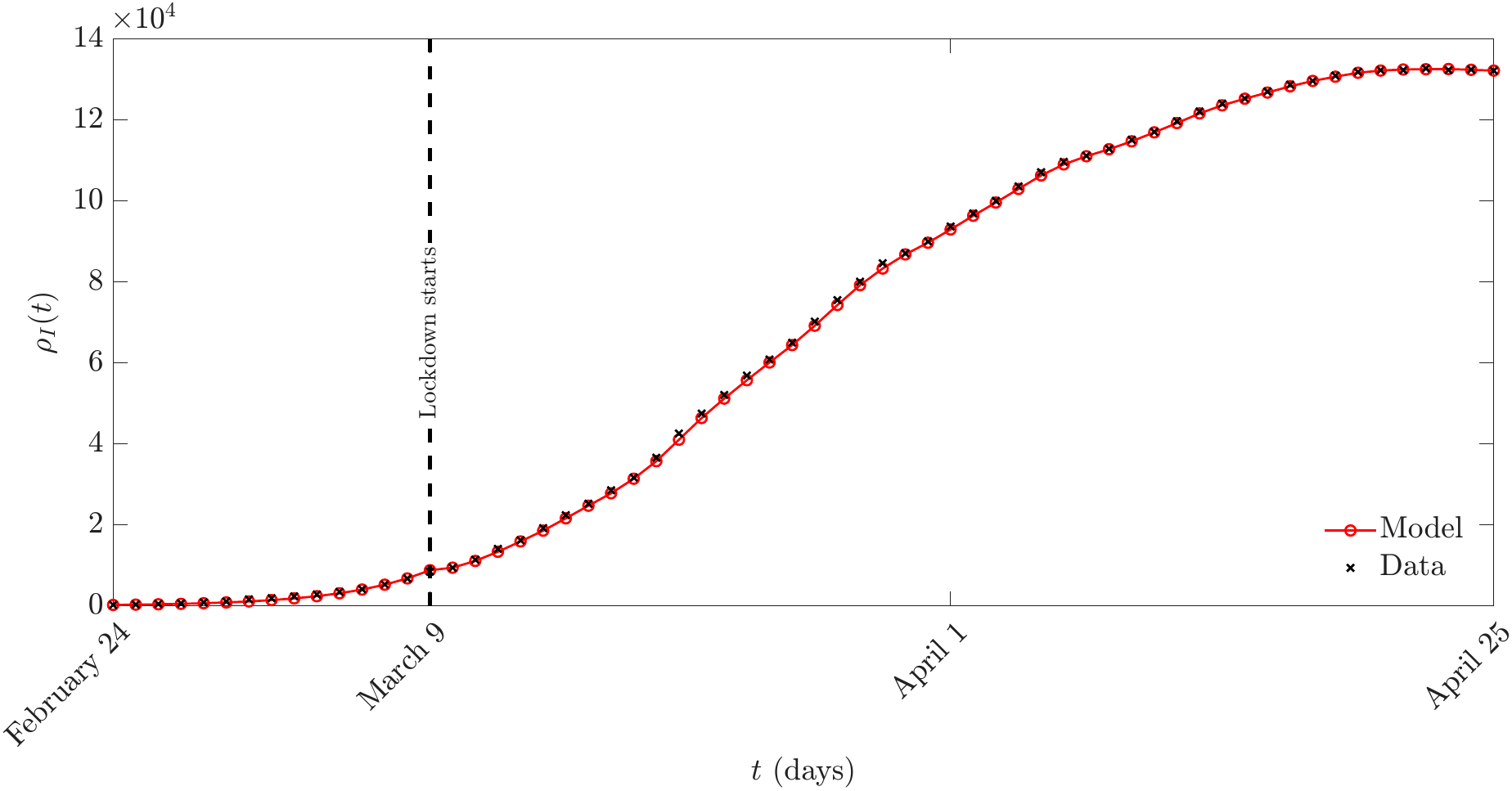}
\caption{\textbf{Model predictions with optimized $\bm{\mu(\rho_I(t))}$.} Time variation of the calibrated parameter $\mu(\rho_I(t))$ as predicted by the minimization problem \eqref{Minimization2} (left) and comparison between the data and their approximation provided by the model for the number of infected $\rho_I(t)$ in Italy, from February 24 to April 25, 2020 (right).}
\label{fig:Predictions-Whole-muI}
\end{figure}

To couple the epidemiological data with opinion data, we determine the values of $\mu(\rho_I(t))$ by solving the optimization problem for the model \eqref{eq:ffe macro 0}--\eqref{eq:cte macro 1} simulated in the time span $t \in [t_L+1, t_f]$ where $t_L+1$ and $t_f$ correspond respectively to the first day after lockdown started (March 10, 2020) and the last day of the observation window (April 25, 2020), and for a sequence of time steps $t_i$ (days) over a moving time window of three days, namely averaging the fitting over three days. This strategy is needed in order to correctly assess the evolution of the current risk perception, depending on the containment measures that are in place. More precisely, we solve the constrained least-square minimization problem
\begin{equation} \label{Minimization2}
\min_{\mu(\rho_I(t_i)) \in [-1,1]} \mathcal C(\rho_I,\rho_R, \hat \rho_I, \hat \rho_R) 
\end{equation}
with cost functional 
\begin{equation}
\label{eq:cost}
\begin{split}
\mathcal C(\rho_I, &\rho_R, \hat \rho_I, \hat \rho_R) = \\[2mm]
&\quad(1-\eta) \frac{\norm{ \rho_I(t) - \hat{\rho}_I(t) }_{L^2([t_i-1,t_i+2])}}{\norm{ \hat{\rho}_I(t) }_{L^2([t_i-1,t_i+2])}} + \eta \frac{\norm{ \rho_I(t) + \rho_R(t) - \hat{\rho}_I(t) - \hat{\rho}_R(t) }_{L^2([t_i-1,t_i+2])}}{\norm{ \hat{\rho}_I(t) + \hat{\rho}_R(t) }_{L^2([t_i-1,t_i+2])}},
\end{split}\end{equation}
while the parameters $\beta, \gamma$ are taken from the pre-lockdown analysis, and we fix $\eta = 10^{-3}$. The evolution over time of the optimal $\mu(\rho_I(t))$ and the resulting predictions obtained by running our macroscopic SIR model \eqref{eq:ffe macro 0}--\eqref{eq:cte macro 1} with these optimized values are displayed in Figure \ref{fig:Predictions-Whole-muI}.

As a natural follow-up question, we are now interested in determining the optimal trend of the calibrated parameters $\mu(\rho_I(t))$ that is in line with the theoretical form \eqref{eq:mu} proposed in \cite{BTZ}. In particular, notice that our choice of fixing $\mu(\rho_I) \equiv -1$ in the pre-lockdown phase also came from the fact that, in view of recovering a shape like \eqref{eq:mu}, it is natural to start from the value $-1$ when there are no infected, i.e. when $\rho_I(t_0) = 0$.

\begin{figure}[h!]
\centering
\includegraphics[width = 0.6\linewidth]{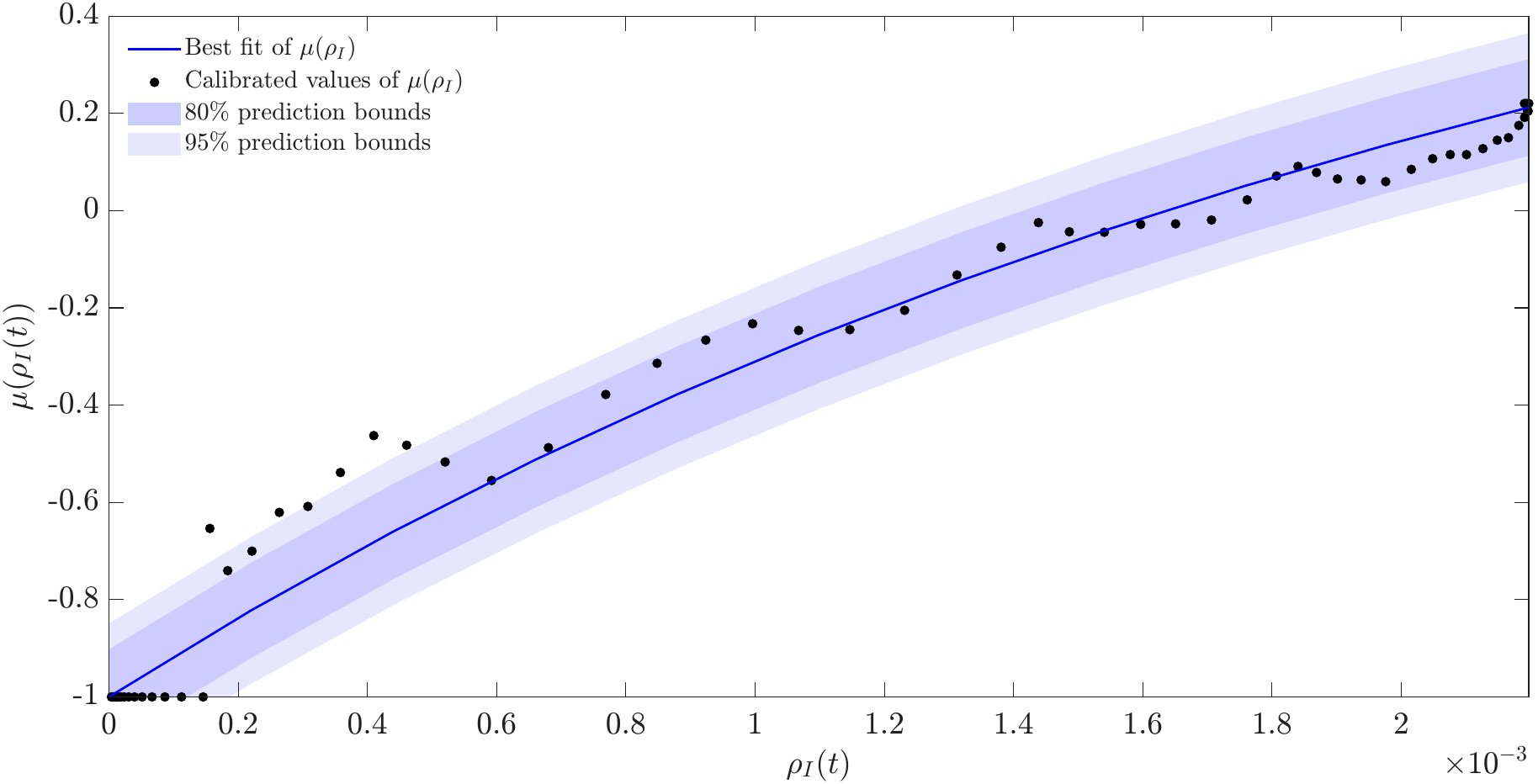}
\caption{\textbf{Optimal shape of $\bm{\mu(\rho_I(t))}$.} Best fit for the calibrated parameters $\mu(\rho_I(t))$, with respect to the number of infected individuals $\rho_I(t)$. We recover the function \eqref{eq:mu} with an exponent $r \approx 4.77 \cdot 10^2$.}
\label{fig:Fit-muI}
\end{figure}

Starting from here, our aim is to find an optimal value for the parameter $r > 0$ characterizing the intensity of risk perception in the model function \eqref{eq:mu}. This can be done by determining the best fit for the parameters $\mu(\rho_I(t))$, as functions of the mass fraction of infected individuals $\rho_I(t)$. We find that the theoretical shape given by \eqref{eq:mu} is recovered for an exponent $r \approx 4.23 \times 10^2$, with a $95\%$ confidence interval $[4.08, 4.38] \times 10^2$ and a coefficient of determination $R^2 \approx 0.97$. The corresponding fit is shown in Figure \ref{fig:Fit-muI}, along with the associated $80\%$ and $95\%$ prediction bounds.

The estimated value of the exponent $r$ suggests that, during the first wave of the epidemic, changes in collective risk perception were highly sensitive to variations in the number of infected individuals. This indicates that behavioral adaptation was not gradual, but rather accelerated once the epidemic reached a sufficiently high incidence.

\subsubsection{Quantifying the contribution of social media analysis}

As a final application, we investigate the extent to which the sentiment extracted from the selected Twitter sample contributes to the behavioral opinion variable governing the epidemic dynamics. Denoting by $\hat{m}(t)$ the average sentiment measured from the Twitter data, we infer the weight that this sample should receive in the population average opinion in order to reproduce the observed epidemic dynamics.

\begin{figure}[h!]
\centering
\includegraphics[width = 0.48\linewidth]{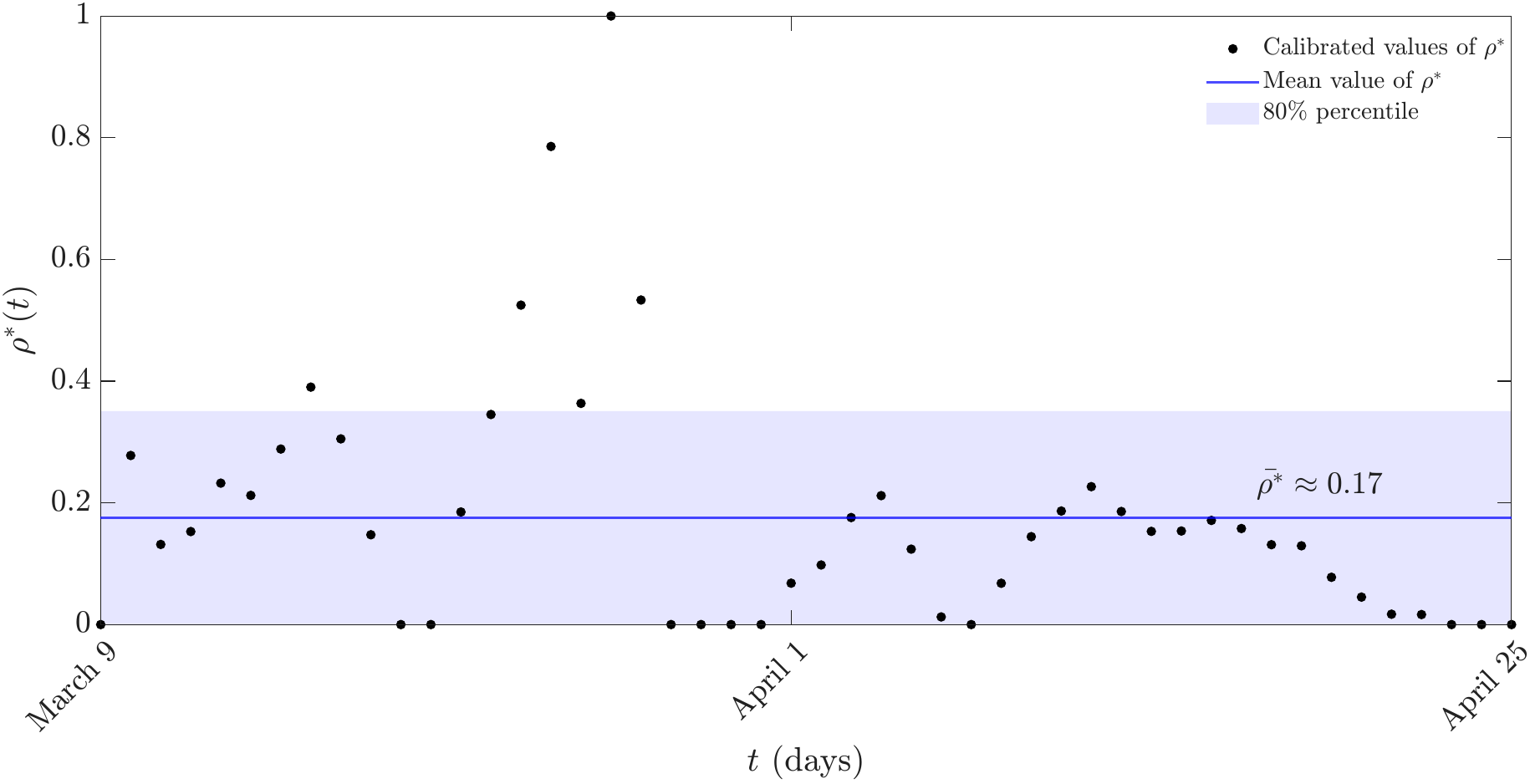} \hfill
\includegraphics[width = 0.48\linewidth]{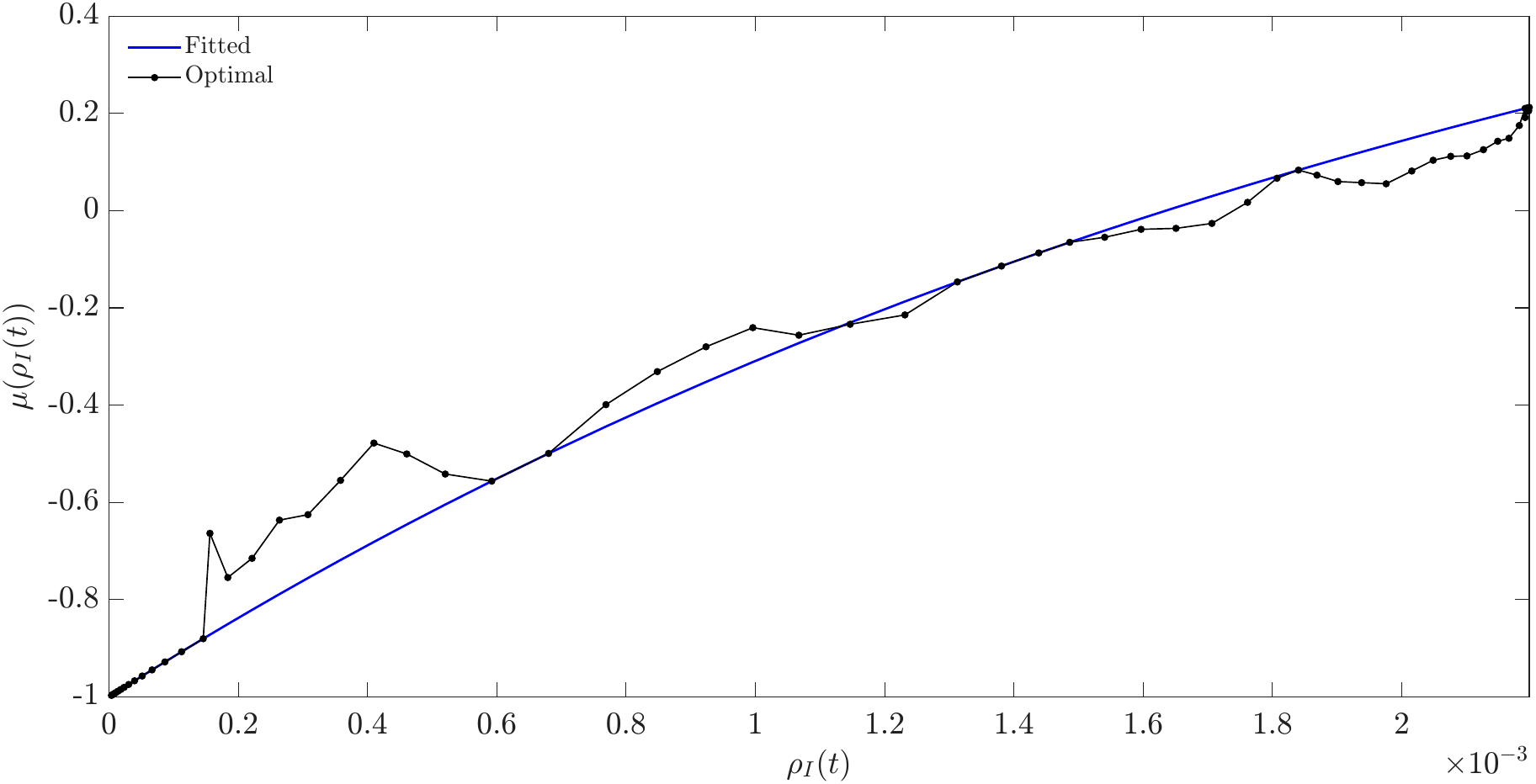}
\caption{\textbf{Analysis of the Twitter sample.} Time variation of the optimal mass fraction $\rho^*(t)$ as predicted by the minimization problem \eqref{Minimization3} (left), and comparison between the theoretical fit $\mu^{\mathrm{fit}}(\rho_I(t))$ for the calibrated parameters $\mu(\rho_I(t))$ and the optimal values of $m(t)$ computed from \eqref{eq:mu-opt} (right).}
\label{fig:Fit-p}
\end{figure}

We start with the observation that the specific regime $\tau \ll 1$ that we are considering implies that the total mean opinion $m(t)$ has formally relaxed toward the risk perception $\mu(\rho_I(t))$ for any fixed $t > 0$, hence the evolution of the model \eqref{eq:cte macro 0}--\eqref{eq:cte macro 1} follows that of the limit system \eqref{eq:cte limit}. Therefore, using the proposed fit $\mu^{\mathrm{fit}}(\rho_I(t)) = 1 - 2(1 - \rho_I(t))^r$ (with $r$ given in Table \ref{tab:inferred-parameters}) for the optimal parameters $\mu(\rho_I(t))$, we argue that the mass fraction $\rho^*(t)$ of the subpopulation associated with the mean opinion $\hat{m}(t)$ of the considered Twitter sample can be determined by optimizing a suitable convex combination of the form
\begin{equation} \label{eq:mu-opt}
m(t) \underset{\tau \ll 1}{\sim} (1-\rho^*(t)) \mu^{\mathrm{fit}}(\rho_I(t)) + \rho^*(t) \hat{m}(t),
\end{equation}
to match the evolution of the infected individuals for any time $t \in [t_L+1,t_f]$ ranging in the time span that goes from March 10 to April 25, 2020. To do this, we follow the previous calibration procedure and solve, in the considered temporal window of observation and for a sequence of three-day intervals around daily time steps $t_i$, the least-square minimization problem
\begin{equation} \label{Minimization3}
\min_{\rho^*(t_i) \in [0,1]} \mathcal C(\rho_I,\rho_R,\hat \rho_I,\hat \rho_R),
\end{equation}
where the cost functional has been defined in \eqref{eq:cost} and now $\mu(\rho_I(t))$ has the explicit form $\mu^{\mathrm{fit}}(\rho_I(t))$ displayed in Figure \ref{fig:Fit-muI}, and the parameter to calibrate is now the mass fraction $\rho^*(t) \in [0,1]$. Notice that we still fix $\beta$ and $\gamma$ to the values determined by the pre-lockdown analysis, provided in Table \ref{tab:inferred-parameters}. The evolution over time of the computed values $\rho^*(t)$ and the relative distance between the proposed fit $\mu^{\mathrm{fit}}(\rho_I(t))$ and the optimal values of $m(t)$ from \eqref{eq:mu-opt} are shown in Figure \ref{fig:Fit-p}. 

In the proposed mixture representation, the calibrated weight assigned to Twitter sentiment has a temporal average of approximately $17\%$. This value quantifies the effective contribution of the Twitter-based indicator to the opinion variable required to reproduce the observed epidemic dynamics. This value is a model-dependent measure of the explanatory contribution of Twitter sentiment relative to the epidemic-driven risk-perception component.

In particular, the calibrated values of $\rho^*(t)$ indicate that the sentiment extracted from Twitter contributes only partially to the effective opinion governing the epidemic dynamics. This finding is consistent with the fact that social media users represent a specific and potentially biased subset of the population. Nevertheless, the estimated contribution is sufficiently large to suggest that online sentiment captures a meaningful fraction of the behavioral response driving epidemic evolution. Therefore, despite its intrinsic sampling bias, social media information may provide valuable real-time evidence for monitoring behavioral changes when interpreted through an appropriate mathematical model.

\section*{Discussion}

We introduced a multiscale kinetic framework coupling opinion formation and epidemic dynamics, derived its Fokker–Planck approximation, and obtained closed macroscopic equations through a suitable equilibrium closure. We then combined this mathematical framework with epidemiological observations and sentiment scores extracted from Italian Twitter posts through a machine-learning-based sentiment analysis, allowing behavioral information to be incorporated into the model calibration.

The findings suggest that an endogenous behavioral response is compatible with part of the temporal variation in transmission that would otherwise be represented through externally specified time-dependent transmission parameters. Within the proposed framework, increases in the observed epidemic burden influence the propensity to adopt protective behaviors, which in turn modifies effective transmission. The model therefore provides a representation of the feedback between epidemic evolution and behavioral adaptation.

This interpretation is consistent with empirical evidence showing that changes in social-contact patterns and individual behavior can substantially influence epidemic transmission \cite{mossong, eames}. However, while these studies used observed contacts or behavioral data to characterize changes in transmission, the present framework represents behavioral adaptation as an endogenous process that evolves in response to epidemic burden. 

Indeed, the proposed model extends the classical SIR by allowing transmission to depend on the evolving propensity of the population to adopt protective behaviors.

The analysis also provides an estimate of the contribution of social media platform to the latent opinion variable. Although Twitter data do not represent the entire population, our analysis shows that, when combined with epidemiological observations through the proposed model, they contain valuable information on behavioral adaptation and can be exploited to improve epidemic inference.
Rather than prescribing this variability empirically, the proposed framework relates it to changes in collective opinion induced by the evolving epidemic. Furthermore, the agreement between the inferred risk perception function and the sentiment extracted from social media data supports the relevance of online behavioral indicators for describing population responses during an epidemic.

These findings should be interpreted in light of some limitations. The risk-perception function does not represent a direct measurement of individual risk perception or adherence to protective behavior. Moreover, the study period included the introduction of the national lockdown and simultaneous changes in mobility, social contacts, testing practices and public-health communication. Consequently, the behavioral component inferred by the model may capture both spontaneous behavioral adaptation and policy-induced reductions in contact, whose separate contributions cannot be clearly identified from the available data. Reported active cases during the first epidemic wave were also affected by varying testing availability, diagnostic practices, reporting delays and the administrative definition of recovery, impacting on data quality and generalizability.  

Overall, the proposed approach provides a flexible mathematical framework for integrating machine-learning-based sentiment analysis with mathematical epidemiology, opening new perspectives for incorporating behavioral data into epidemic forecasting and public-health decision making. Its application to the first COVID-19 wave in Italy illustrates how epidemiological and social-media data can be integrated to investigate feedback between epidemic burden and attitudes to adhere to protective behaviors. Although the inferred behavioral component cannot be disentangled from concurrent policy and surveillance changes, the framework provides a basis for incorporating heterogeneous behavioral information into epidemic modeling.

Future developments should consider richer epidemiological compartmentalizations, heterogeneous contact structures, multiple information sources, including vaccination status and misinformation dynamics data, also integrating social-media with mobility data.

\section*{Acknowledgments}

This work has been written within the activities of the GNFM group of INdAM (National Institute of High Mathematics). A.B. acknowledges the support from the European Union’s Horizon Europe research and innovation programme under the Marie Skłodowska-Curie grant agreement 101110920, project MesoCroMo (A Mesoscopic approach to Cross-diffusion Modelling in population dynamics). M.Z. acknowledges partial support from the PRIN2022PNRR project No.P2022Z7ZAJ, European Union-NextGenerationEU. M.Z. acknowledges partial support by ICSC - Centro Nazionale di Ricerca in High Performance Computing, Big Data and Quantum Computing, funded by European Union-NextGenerationEU.

\section*{Data availability statement}

The data that support the findings of this study are available from the corresponding author upon reasonable request.


\nocite{*}
\bibliographystyle{plain}
\bibliography{Bibliography_SIR}

\end{document}